\documentclass[a4paper,fleqn,usenatbib, twocolumn]{aastex63}

\usepackage[T1]{fontenc}
\usepackage{ae,aecompl}

\usepackage{graphicx}	
\usepackage{url}        
\usepackage{multirow}   
\usepackage{amsmath}	
\usepackage{amssymb}	
\usepackage{textcomp}   
\usepackage{hyperref}   
\usepackage{booktabs}
\usepackage{dashrule}

\usepackage{savesym}
\savesymbol{tablenum}
\usepackage{siunitx}
\restoresymbol{SIX}{tablenum}

\accepted{to ApJS; September 3, 2021}

\shorttitle{The HARP and SCUBA-2 High Resolution Terahertz Andromeda Galaxy Survey - HASHTAG}
\shortauthors{M. W. L. Smith}

\begin{document}

\title{The HASHTAG project: The First Submillimeter Images of the Andromeda Galaxy from the Ground.}

\correspondingauthor{Matthew W. L. Smith}
\email{Matthew.Smith@astro.cf.ac.uk}

\author[0000-0002-3532-6970]{Matthew W. L. Smith}
  \affiliation{School of Physics \& Astronomy, Cardiff University, The Parade, Cardiff, CF24 3AA, UK.}

\author[0000-0002-7394-426X]{Stephen A. Eales}
  \affiliation{School of Physics \& Astronomy, Cardiff University, The Parade, Cardiff, CF24 3AA, UK.}

\author[0000-0002-0012-2142]{Thomas G. Williams}
  \affiliation{Max-Planck-Institut f\"{u}r Astronomie, K\"{o}nigstuhl 17, D-69117, Heidelberg, Germany.}

\author[0000-0002-3810-1806]{Bumhyun Lee}
  \affiliation{Kavli Institute for Astronomy and Astrophysics, Peking University, Beijing 100871, China.}
  \affiliation{Department of Astronomy, Yonsei University, 50 Yonsei-ro, Seodaemun-gu, Seoul, 03722, South Korea}

\author[0000-0002-7172-6306]{Zongnan Li}
  \affiliation{School of Astronomy and Space Science, Nanjing University, Nanjing 210023, China.}
  \affiliation{Key Laboratory of Modern Astronomy and Astrophysics, Nanjing University, Nanjing 210023, China.}

\author[0000-0003-2767-0090]{Pauline Barmby}
  \affiliation{Department of Physics \& Astronomy, Institute for Earth \& Space Exploration, University of Western Ontario, 1151 Richmond Street, London, Ontario, Canada.}

\author[0000-0003-4980-1012]{Martin Bureau}
  \affiliation{Sub-department of Astrophysics, Department of Physics, University of Oxford, Denys Wilkinson Building, Keble Road, Oxford, OX1 3RH, UK.}

\author{Scott Chapman}
  \affiliation{Dalhousie University, Halifax NS, Canada}

\author{Brian S. Cho}
  \affiliation{Department of Physics and Astronomy, Seoul National University, Gwanak-gu, Seoul 08826, Republic of Korea.}

\author[0000-0003-1440-8552]{Aeree Chung}
  \affiliation{Department of Astronomy, Yonsei University, 50 Yonsei-ro, Seodaemun-gu, Seoul, 03722, South Korea.}

\author[0000-0003-0014-1527]{Eun Jung Chung}
  \affiliation{Department of Astronomy and Space Science, Chungnam National University, 99 Daehak-ro, Yuseong-gu, Daejeon 34134, Republic of Korea.}

\author{Hui-Hsuan Chung}
  \affiliation{Institute of Astronomy, National Tsing Hua University, Hsinchu, 30013, Taiwan.}

\author[0000-0001-7959-4902]{Christopher J. R. Clark}
  \affiliation{Space Telescope Science Institute, 3700 San Martin Drive, Baltimore, Maryland 21218, USA.}

\author[	0000-0002-9548-5033]{David L. Clements}
  \affiliation{Astrophysics Group, Physics Department, Blackett Laboratory, Imperial College London, Prince Consort Road, London SW7 2AZ, UK.}

\author[0000-0003-4932-9379]{Timothy A. Davis}
  \affiliation{School of Physics \& Astronomy, Cardiff University, The Parade, Cardiff, CF24 3AA, UK.}

\author[0000-0001-9419-6355]{Ilse De Looze}
  \affiliation{Sterrenkundig Observatorium, Ghent University, Krijgslaan 281 – S9, B-9000 Ghent, Belgium.}
  \affiliation{Department of Physics and Astronomy, University College London, London WC1E 6BT, UK.}

\author[0000-0002-5881-3229]{David J. Eden}
  \affiliation{Astrophysics Research Institute, Liverpool John Moores University, IC2, Liverpool Science Park, 146 Brownlow Hill, Liverpool, L3 5RF, UK.}

\author[0000-0002-7301-3879]{Gayathri Athikkat-Eknath}
  \affiliation{School of Physics \& Astronomy, Cardiff University, The Parade, Cardiff, CF24 3AA, UK.}

\author{George P. Ford}
  \affiliation{School of Physics \& Astronomy, Cardiff University, The Parade, Cardiff, CF24 3AA, UK.}

\author[0000-0003-0007-2197]{Yu Gao}
  \affiliation{Department of Astronomy, Xiamen University, Xiamen, Fujian, 361005, People's Republic of China.}
  \affiliation{Purple Mountain Observatory \& Key Laboratory for Radio Astronomy, Chinese Academy of Sciences, 10 Yuanhua Road, Nanjing 210023, People’s Republic of China.}
  
\author[0000-0001-6789-6196]{Walter Gear}
  \affiliation{Centre for Astronomy, Department of Physics, National University of Ireland Galway, University Road, Galway H91 TK33, Ireland.}

\author[0000-0003-3398-0052]{Haley L. Gomez}
  \affiliation{School of Physics \& Astronomy, Cardiff University, The Parade, Cardiff, CF24 3AA, UK.}

\author[0000-0002-7203-5996]{Richard de Grijs}
  \affiliation{Department of Physics \& Astronomy, Macquarie University, Balaclava Road, Sydney, NSW 2109, Australia.}
  \affiliation{Research Centre for Astronomy, Astrophysics and Astrophotonics, Macquarie University, Balaclava Road, Sydney, NSW 2109, Australia.}

\author[0000-0002-3938-4393]{Jinhua He}
  \affiliation{Yunnan Observatories, Chinese Academy of Sciences, 396 Yangfangwang, Guandu District, Kunming, 650216, P. R. China.}
  \affiliation{Chinese Academy of Sciences South America Center for Astronomy, National Astronomical Observatories, CAS, Beijing 100101, China.}
  \affiliation{Departamento de Astronom\'{i}a, Universidad de Chile, Casilla 36-D, Santiago, Chile.}

\author[0000-0001-6947-5846]{Luis C. Ho}
  \affiliation{Kavli Institute for Astronomy and Astrophysics, Peking University, Beijing 100871, China.}
  \affiliation{Department of Astronomy, School of Physics, Peking University, Beijing 100871, China}

\author[0000-0002-3055-3154]{Thomas M. Hughes}
  \affiliation{Chinese Academy of Sciences South America Center for Astronomy, National Astronomical Observatories, CAS, Beijing 100101, China.}
  \affiliation{Instituto de F\'{i}sica y Astronom\'{i}a, Universidad de Valpara\'{i}so, Avda. Gran Breta\~{n}a 1111, Valpara\'{i}so, Chile.}
  \affiliation{CAS Key Laboratory for Research in Galaxies and Cosmology, Department of Astronomy, University of Science and Technology of China, Hefei 230026, China.}
  \affiliation{School of Astronomy and Space Science, University of Science and
Technology of China, Hefei 230026, China.}

\author{Sihan Jiao}
  \affiliation{National Astronomical Observatories, Chinese Academy of Sciences, 20A Datun Road, Chaoyang District, Beijing 100012, China.}
  \affiliation{University of Chinese Academy of Sciences, 100049, Beijing, PR China.}

\author[0000-0003-0355-6437]{Zhiyuan Li}
  \affiliation{School of Astronomy and Space Science, Nanjing University, Nanjing 210023, China.}
  \affiliation{Key Laboratory of Modern Astronomy and Astrophysics, Nanjing University, Nanjing 210023, China.}

\author[0000-0003-2743-8240]{Francisca Kemper}
  \affiliation{European Southern Observatory, Karl-Schwarzschild-Str. 2, 85748 Garching, Germany.}
  \affiliation{Institute of Astronomy and Astrophysics, Academia Sinica, No. 1, Sec. 4, Roosevelt Rd., Taipei 10617, Taiwan.}

\author[0000-0002-3036-0184]{Florian Kirchschlager}
  \affiliation{Department of Physics and Astronomy, University College London, London WC1E 6BT, UK.}

\author[0000-0001-9605-780X]{Eric W. Koch}
  \affiliation{Center for Astrophysics | Harvard \& Smithsonian, 60 Garden St., 02138 Cambridge, MA, USA.}

\author[0000-0002-5105-344X]{Albert K. H. Kong}
  \affiliation{Institute of Astronomy, National Tsing Hua University, Hsinchu, 30013, Taiwan.}

\author[0000-0003-1700-5740]{Chien-Hsiu Lee}
  \affiliation{NSF's National Optical-Infrared Astronomy Research Laboratory, Tucson, AZ 85742, USA.}

\author[0000-0002-0030-8051]{En-Tzu Lin}
  \affiliation{Institute of Astronomy, National Tsing Hua University, Hsinchu, 30013, Taiwan.}

\author[0000-0002-6956-0730]{Steve Mairs}
  \affiliation{East Asian Observatory, 660 N. A'oh\={o}k\={u} Place, University Park, Hilo, HI 96720, USA.}

\author[0000-0001-9033-4140]{Micha{\l} J. Micha{\l}owski}
  \affiliation{Astronomical Observatory Institute, Faculty of Physics, Adam Mickiewicz University, ul.~S{\l}oneczna 36, 60-286 Pozna{\'n}, Poland}

\author[0000-0002-8557-3582]{Kate Pattle}
  \affiliation{Centre for Astronomy, Department of Physics, National University of Ireland Galway, University Road, Galway H91 TK33, Ireland.}

\author{Yingjie Peng}
  \affiliation{Kavli Institute for Astronomy and Astrophysics, Peking University, Beijing 100871, China.}

\author[0000-0003-4164-5588]{Sarah E. Ragan}
  \affiliation{School of Physics \& Astronomy, Cardiff University, The Parade, Cardiff, CF24 3AA, UK.}

\author[0000-0002-6529-202X]{Mark G. Rawlings}
  \affiliation{East Asian Observatory, 660 N. A'oh\={o}k\={u} Place, University Park, Hilo, HI 96720, USA.}

\author[0000-0001-6854-7545]{Dimitra Rigopoulou}
  \affiliation{Sub-department of Astrophysics, Department of Physics, University of Oxford, Denys Wilkinson Building, Keble Road, Oxford, OX1 3RH, UK.}

\author[0000-0003-4357-3450]{Amelie Saintonge}
  \affiliation{Department of Physics and Astronomy, University College London, London WC1E 6BT, UK.}

\author{Andreas Schruba}
  \affiliation{Max-Planck-Institut f\"{u}r extraterrestrische Physik, Giessenbachstra{\ss}e~1, D-85748 Garching, Germany.}

\author[0000-0002-4154-4309]{Xindi Tang}
  \affiliation{Xinjiang Astronomical Observatory, Chinese Academy of Sciences, 830011 Urumqi, P. R. China.}

\author[0000-0003-4874-0369]{Junfeng Wang}
  \affiliation{Department of Astronomy, Xiamen University, Xiamen, Fujian, 361005, People's Republic of China.}

\author{Anthony P. Whitworth}
  \affiliation{School of Physics \& Astronomy, Cardiff University, The Parade, Cardiff, CF24 3AA, UK.}

\author[0000-0001-5817-0991]{Christine D. Wilson}
  \affiliation{Department of Physics and Astronomy, McMaster University, 1280 Main St W, Hamilton, ON, L8S 4M1 Canada.}

\author[0000-0002-3426-5854]{Kijeong Yim}
  \affiliation{Department of Astronomy and Space Science, Chungnam National University, 99 Daehak-ro, Yuseong-gu, Daejeon 34134, Republic of Korea.}

\author{Ming Zhu}
  \affiliation{National Astronomical Observatories, Chinese Academy of Sciences, 20A Datun Road, Chaoyang District, Beijing 100012, China.}


\begin{abstract}

Observing nearby galaxies with submillimeter telescopes on the ground has two major challenges.
First, the brightness is significantly reduced
at long submillimeter wavelengths compared to the brightness at the peak of the dust emission. Second,
it is necessary to use a high-pass spatial filter to remove atmospheric noise on large angular scales,
which has the unwelcome by-product of also removing the galaxy's large-scale structure.
We have developed a technique for producing high-resolution submillimeter
images of galaxies of large angular size by using the telescope on the
ground to determine the small-scale structure (the large Fourier components) and
a space telescope (\textit{Herschel} or \textit{Planck}) to determine the large-scale structure (the small
Fourier components). Using this technique, we are carrying out
the HARP and SCUBA-2 High Resolution Terahertz Andromeda Galaxy Survey (HASHTAG),
an international Large Program on the James Clerk Maxwell Telescope, with one
aim being to produce the first high-fidelity high-resolution submillimeter
images of Andromeda. In this paper, we 
describe the survey, the method we have developed for combining the space-based
and ground-based data, and present the first HASHTAG images
of Andromeda at 450 and \SI{850}{\micro\meter}. We also have created a method to predict the CO($J$=3--2)
line flux across M\,31, which contaminates the \SI{850}{\micro\meter} band. We find that while normally
the contamination is below our sensitivity limit, the contamination can be 
significant (up to 28\%) in a few of the brightest regions of the 10\,kpc ring.
We therefore also provide images with the predicted line emission removed.

\end{abstract}

\keywords{galaxies: individual (M31) - galaxies: ISM - galaxies: star formation - submillimeter: galaxies - methods: observations - submillimeter: ISM}



\section{Introduction}

The Andromeda Galaxy (Messier\,31) is possibly the most frequently observed galaxy in the
sky. Galaxies in the Local Group are important for the
obvious reason that they are closest, allowing us to study galaxies
in the greatest possible detail, but they are
also important because they are the only galaxies in which we can detect large
numbers of individual stars. The ability to see stars adds a large number of
investigative tools to the astronomer's toolkit, which are not possible to
use on galaxies outside the Local Group.

There are only three spiral galaxies in the Local Group: our own, Andromeda, and the Triangulum (Messier\,33). The Triangulum has a mass roughly ten times less
than our own, but Andromeda has a mass and other properties that are quite similar to our own \citep{yin2009}.
However, there are also some interesting differences. Andromeda has a larger bulge \citep{yin2009}, less obvious spiral arms \citep{gordon2006,kirk2015},
and much of the star formation in the galaxy is occurring in a large ring \citep{ford2013}. The cause of this ring is unknown. One interesting suggestion is that the ring may be the result of the dwarf galaxy M\,32 passing through the center of the disk, generating a density wave, and thus a wave of star birth that propagates outwards through the disk \citep{block2006}.
This now seems unlikely since the star-formation history in the disk has no obvious radial gradient \citep{lewis2015}, and the cause of the ring remains a mystery.

An iconic naked-eye object, Andromeda has now been observed by professional astronomers for
over a thousand years. It was discovered, or at least first mentioned, in 964 by the Persian
astronomer Abd al Rahman al-Sufi ({\it Book of Fixed Stars}). In the eighteenth 
century, it was observed by William Herschel, who noticed the red colors of the
central bulge \citep{herschel1785}. In the twentieth century, Edwin Hubble used the
Cepheid variables in the Andromeda Nebula to show that the
nebula is actually a galaxy \citep{hubble1929}. In the modern era, Andromeda has been surveyed by virtually every modern observatory. A very incomplete list of the telescopes that have surveyed Andromeda includes XMM-Newton which surveyed the galaxy in X-ray \citep{Stiele2011}, GALEX in the ultraviolet \citep[UV, ][]{Thilker2005}, {\it Spitzer} in the mid- and far-infrared \citep{barmby2005,gordon2006}, {\it Herschel} in the far-infrared and submillimeter \citep{fritz2012,smith2012,draine2014}, Westerbork \& VLA in the radio 21\,cm line  \citep{braun2009, Koch2021}, and the IRAM 30 metre telescope in the CO($J$=1--0) line \citep{Nieten2006}. The northern third of the galaxy has also been observed with {\it Hubble} in the Panchromatic Hubble Andromeda Treasury Survey (PHAT), which
detected $\simeq$117 million stars \citep{dalcanton2012}.

A glaring omission on the list is a submillimeter telescope on the ground,
from where it is possible to get much better angular resolution than is possible with the small mirrors of space telescopes. The angular resolution of the {\it Herschel} observations of Andromeda
at \SI{500}{\micro \meter} was 36 arcsec (FWHM), which is equivalent at the distance of Andromeda
\citep[780\,kpc, ][]{deGrijs2014} to a spatial resolution of about 136\,pc, roughly the size of
an association of GMCs. But with the SCUBA-2 camera on the James Clerk Maxwell Telescope (JCMT), the world's largest submm telescope, it should be possible
to map Andromeda at \SI{450}{\micro\meter} with a resolution of $\simeq$8 arcsec, equivalent
to a spatial resolution of $\simeq$30\,pc, slightly less than the size
of a typical GMC. The reason that such a map has not previously been created is due to the atmospheric
noise in the sub-millimeter waveband, which requires the data to be filtered strongly
on angular sizes larger than that of the field-of-view of the camera, which is 45 arcmin$^2$ in the
case of SCUBA-2 \citep{holland2013}, much smaller than the $\rm 3 \times 1$ deg$^2$ ($\rm \sim10^4\,arcmin^2$) that Andromeda occupies on the sky.

The solution to the problem is to combine data from a space
observatory with data from a camera on the ground, using the camera on the
ground to produce the high-resolution information and the observatory
in space to determine the large-scale structure. In Fourier terms, we use the
space data, from {\it Herschel} and {\it Planck} to provide the low-\textit{k} Fourier 
components and the camera on the ground to provide the high-\textit{k} components.

We are using this technique to carry out a large survey of Andromeda with the
JCMT: The HARP and SCUBA-2 High Resolution Terahertz Andromeda Galaxy Survey (henceforth HASHTAG). HASHTAG has been awarded 276 hours on the telescope, most of
which is being used to carry out a survey with SCUBA-2 at 450 and 
\SI{850}{\micro \meter} (221/275 hours). The rest of the time has been used to carry out a survey in the
CO($J$=3--2) line with HARP in 12 regions covering a total area
of 60\,arcmin$^2$ within Andromeda's disk. 
By combining the data from SCUBA-2 with the
{\it Herschel} images of Andromeda at six wavelengths, using an algorithm that does not
require any smoothing of the data or assumptions about the temperature of the
dust \citep{marsh2015}, our goal is to produce maps of the bolometric dust emission and of the
dust column density as a function of dust temperature and dust emissivity index ($\beta$) with a resolution
of $\simeq$25\,pc at 
$\simeq$70,000 independent
positions within the galaxy, maps which will be used for a large range of scientific projects.

\begin{figure*}
  \centering
  \includegraphics[trim=0mm 0mm 0mm 0mm, clip=True, width=0.95\textwidth]{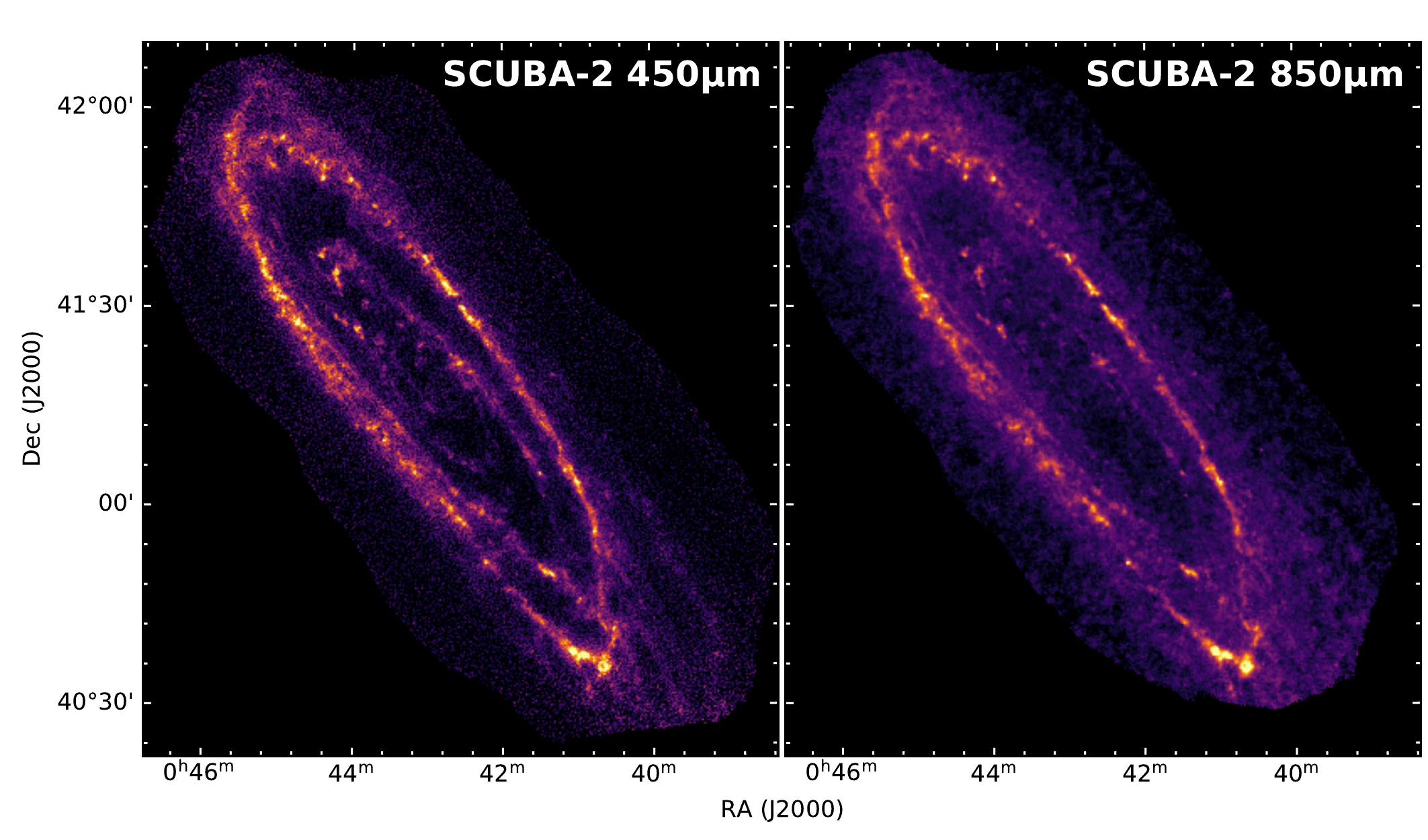}
  \caption{The HASHTAG images created from the first $\sim$70\% of the final SCUBA-2 dataset. The 450 and \SI{850}{\micro\meter} images
  have been smoothed with a 7.9, and 13\arcsec\ FWHM Gaussian, respectively. For the raw-resolution images see Figure~\ref{fig:DR1images}.}
  \label{fig:smoDR1images}
\end{figure*}

The CO part of HASHTAG has been completed and the results published in
\citet{Li2020}. The continuum part of HASHTAG is now about 70\% complete
and we have recently made the first full mosaics. The images cover the entire galaxy
and have reached full sensitivity in the one third of the
disk that has been covered by {\it Hubble} by PHAT \citep{dalcanton2012}, by the Combined Array for Research
in Millimeter wave Astronomy in the CO($J$=1--0) line \citep{CalduPrimo2016}, and higher-resolution ($\sim$10\arcsec) VLA H{\sc i} observations \citep{Koch2021}. 
By using the {\it Herschel} image
at \SI{500}{\micro \meter} \citep{fritz2012} and the {\it Planck} image at \SI{850}{\micro\meter} \citep{planck2015} to
fill in the low-frequency (low-\textit{k}) Fourier components, we have produced the first high-fidelity
images of Andromeda from the ground, at two wavelengths, 450 and \SI{850}{\micro\meter} (Figure~\ref{fig:smoDR1images}). These images are
being used in the first round of HASHTAG science papers.

This paper gives an overview of HASHTAG and describes the observations and
data reduction procedure used to generate the images shown in Figure~\ref{fig:smoDR1images},
including a description of the technique we have developed to combine the
space-based and ground-based continuum submillimeter observations, and the measures
we have taken to optimize the pipeline parameters. 
All data products and codes presented here are available on the HASHTAG website\footnote{\url{https://hashtag.astro.cf.ac.uk/} \label{foot:web}}.
Future data releases will also be made available on this site.

The organisation of the paper is as follows. 
Section~\ref{sec:scienceOverview} gives an overview of the science programme that can be carried out
with these images.
Section~\ref{sec:surveyStragey} describes the observing
method and mapping strategy. Section~\ref{sec:dataReduc} describes the data-reduction pipeline and
the method used to combine the space-based and ground-based
data. Sections~\ref{sec:simulations} and \ref{sec:450sim} describes the extensive simulations that we have carried out
to optimize and test the data-reduction pipeline and combination method
described in Section~\ref{sec:dataReduc}. Section~\ref{sec:realData} presents our final reduced maps, including some simple analysis
of their properties. Finally, Section~\ref{sec:CO} describes how we estimate the contamination from CO($J$=3--2) in our
continuum observations.

\section{Overview of Science Programme}
\label{sec:scienceOverview}

In this section we give a brief overview of the scientific projects that
will be possible with the HASHTAG dataset. These mostly fall into two categories:
dust and star formation.

\subsection{Dust}

Dust itself is interesting for two main reasons. First, it is of great intrinsic
interest because it is a vital phase of the interstellar medium (ISM), containing half the heavy
metals \citep{james2002} and being a catalyst in the networks of chemical reactions in the ISM, including the vital one in which atomic hydrogen is transformed
into molecular hydrogen. Second, mapping the continuum emission from
dust grains is a promising method for both mapping the ISM in galaxies
and estimating the total mass of the ISM \citep{hildebrand1983,eales2012,magdis2012,scoville2014}. The standard
tracer of the molecular phase, the CO molecule,
has many well-known disadvantages \citep{bolatto2013}. There is also now
the more fundamental problem that one third of the molecular gas in the Galaxy appears to contain no
CO \citep{abdo2010,planck2011,pineda2013}, and there are even galaxies
in which the fraction of `CO-dark' gas seems to be close to 100\% \citep{Dunne2021}.
Some of the
advantages of using dust grains rather than CO molecules to trace
the ISM are that the dust emission is optically thin, dust
grains are robust and not liable to be destroyed by starlight, and the relationship
between the gas-to-dust ratio and metallicity seems to be much simpler than between
CO abundance and metallicity \citep{eales2012,sandstrom2013,remyruyer2014}. 
    
The biggest contribution that HASHTAG seems likely to make to our understanding
of the dust itself is to show how the properties of dust vary within an individual
galaxy. The earlier {\it Herschel} observations of Andromeda revealed systematic large-scale spatial variation in the properties of dust. The emission from
interstellar dust follows a modified blackbody ($S_{\nu} \propto B_{\nu}(T_{\rm d}) \nu^{\beta}$). The {\it Herschel} observations revealed
that $\beta$ varies radially within Andromeda's disk \citep{smith2012,draine2014,Whitworth2019}, and radial variation in
$\beta$ has subsequently been found in the Galaxy \citep{planck2014}, M\,33 \citep{tabat2014} and in $\simeq$20 other galaxies \citep{hunt2015}, although
the form of the radial variation varies between galaxies \citep{hunt2015}.
There is also now some
evidence that the global value of $\beta$ varies between
galaxies \citep{lamperti2019}. The cause of the variation must be caused by changes
in the structure, physics or chemistry of the dust, although what the key
changes are is currently unknown, one
clue may be that in Andromeda $\beta$ does not appear to differ between
low-density and high-density gas \citep{Eknath2021}.

HASHTAG will produce measurements of $\beta$ at $\simeq$70,000 positions in
Andromeda's disk, effectively producing a dust atlas for Andromeda. The
{\it Hubble} observations of one third of the disk have produced estimates,
from the optical extinction,
of the column-density of dust with the same spatial resolution that HASHTAG will provide
\citep{dalcanton2012}. The combination of measurements of the emission
properties of dust from submm observations and the absorption properties
of dust from optical observations is a powerful one for testing theoretical
dust models. The emission and absorption properties of dust, derived from {\it Hubble} and {\it Herschel} data,
are already inconsistent with all existing dust models \citep{Whitworth2019}. The combination of the high-resolution measurements of $\beta$ from HASHTAG, the {\it Hubble} dust measurements and the maps of the ISM phases, star formation, chemical abundances and other properties that are available for Andromeda offer at least the prospect
of uncovering the physical/chemical causes of the variation in dust.

The use of the dust emission to trace the ISM in Andromeda offers a number
of interesting possibilities. By comparing the dust emission to the CO and H{\sc i} emission it will be possible to search for CO-dark gas in Andromeda \citep{planck2011,sandstrom2013}. It will also be possible to produce catalogues
of GMCs based on dust emission rather than CO. This technique has already
been applied to the {\it Herschel} observations of Andromeda, producing
a catalogue of 326 clouds with masses between
$\rm10^4\,M_{\odot}$ and $\rm 10^7\,M_{\odot}$ \citep{kirk2015}. A more recent
study suggests that clouds found by the dust method have much lower CO-to-dust ratio
than the clouds found from their CO emission \citep{Eknath2021}, suggesting there
is much more variation in the properties of clouds than one would expect from
a CO-selected catalogue. The clouds in the {\it Herschel} catalogue are probably
associations of GMCs rather than single GMCs, but with the extra resolution of HASHTAG it will be possible for the first time to produce a catalogue of dust-selected clouds that are likely to be GMCs rather than GMC associations.

\subsection{Star Formation}

One of the most important properties to measure within a galaxy is
the star-formation rate, but there is still no gold-standard way
of doing this. There are at least 12 different methods which use
different techniques for tracing the obscured and unobscured star formation \citep{kennicutt2012,Speagle2014,davies2016}, all of which have limitations, and none
is clearly better than the others.

HASHTAG will produce high-resolution maps of the bolometric dust emission, which is a direct measurement of the emission from the obscured OB stars, although it is really
an upper limit since much of the bolometric dust emission is
re-radiated emission from the older stellar population \citep{bendo2012,kennicutt2012,bendo2015,viaene2017, ford2013}. However, since Andromeda is close enough to detect individual stars, there is at least the
possibility of combining {\it Hubble} observations of the unobscured
OB stars and the HASHTAG observations of the emission from the obscured
stars to provide a direct measurement of the star-formation rate, rather
than the ones produced by the current methods, which mostly rely on indirect tracers of the
star formation. 
The PHAT team made a first attempt to do this \citep{lewis2017}, using optical extinction measurements to
correct for obscuration, but their method was unable to account for the OB stars
that are still deep in GMCs and are completely hidden by dust. If it is possible
to correct for the part of the bolometric dust emission that is
re-radiated emission from the older stellar population, HASHTAG will provide estimates of numbers of these missing OB stars.


\section{Survey Strategy}
\label{sec:surveyStragey}

HASHTAG is a JCMT Large Programme (ID: M17BL005), and is split into two components: the continuum submillimeter observations 
of the entire galaxy and observations in the CO($J$=3--2) line of selected regions. 

We used 55.3 hours with HARP \citep{buckle2009} to observe 11 
2\arcmin$\times$2\arcmin\ fields, and one 4\arcmin$\times$4\arcmin\ field. We selected these fields to cover a
range of diverse ISM conditions in M\,31 and to maximize overlap with
useful ancillary data e.g. \textit{Herschel} far-infrared spectroscopy \citep{kapala2015}. Our observations were carried out in 2017 using grade-3 weather 
defined as when the opacity at 225\,GHz ($\tau_{\rm 225\,GHz}$) is between 0.08 and 0.12, and reached a sensitivity of $\simeq15$\,mK (antenna temperature, T$_{\rm a}^*$) with an angular resolution of 15\arcsec\ and spectral resolution of 2.6\,km\,s$^{-1}$. A full discussion of our CO observations, is given in \citet{Li2020}. 
The CO($J$=3--2) line falls within the 
\SI{850}{\micro \meter} continuum filter and so our CO spectroscopic mapping are useful for assessing the
effect of line contamination on the continuum measurements.

The larger component of HASHTAG consists of the continuum observations, which were allocated
221 hours in the less-common grade-2 weather ($0.05 < \tau_{\rm 225\,GHz} < -0.12$) and commenced in 2017 (expected to complete 2021). 
SCUBA-2 observes simultaneously at 450 and \SI{850}{\micro\meter}, producing
images at the two wavelengths with the same field-of-view \citep{holland2013}, and angular resolutions of 7.9 and 13.0\arcsec\ at 
450 and \SI{850}{\micro\meter}, respectively \citep{Dempsey2013}.
Our goal was to observe the entire galaxy at both wavelengths.

Our continuum observing strategy is based on our experience in a smaller project in 2015 (project M15BI082, P.I. Smith) and was effectively HASHTAG's pilot
field. The SCUBA-2 Pong observing mode is used for sources greater  than $\sim$5\arcmin\ in size and can be used to map a circular
region of diameter 15, 30 or 60 arcmin (since the design of HASHTAG a 45 arcmin mode has been added). 
For the pilot we made a long {\it Pong} exposure of a single circular region of diameter
30 arcmin. For this field, we chose the position of the center so that this
circular region would also include a significant area in
which there was no obvious sub-millimeter emission in the
{\it Herschel} images from the galaxy itself, since we knew this `background region' would help
with the data reduction to converge. The maximum integration time for a SCUBA-2 observation
is 45 minutes. We chose an integration time of 43 minutes, repeating it
37 times.
We reached a sensitivity of 44.9 and 3.0\,mJy\,beam$^{-1}$
using 2.0 and 4.0\arcsec\ pixels, at 450 and \SI{850}{\micro\meter}, respectively. 

Given the success of these observations, and our development of a technique
to use space-telescope data to replace the large-scale structure severely suppressed  
by the filtering necessary to remove atmospheric noise (see Sections~\ref{sec:dataReduc} and~\ref{sec:simulations}), we realized it would be practical to use a similar strategy to observe the
whole of Andromeda.

\begin{figure}
  \centering
  \includegraphics[trim=1mm 12mm 1mm 8mm, clip=True, width=0.47\textwidth]{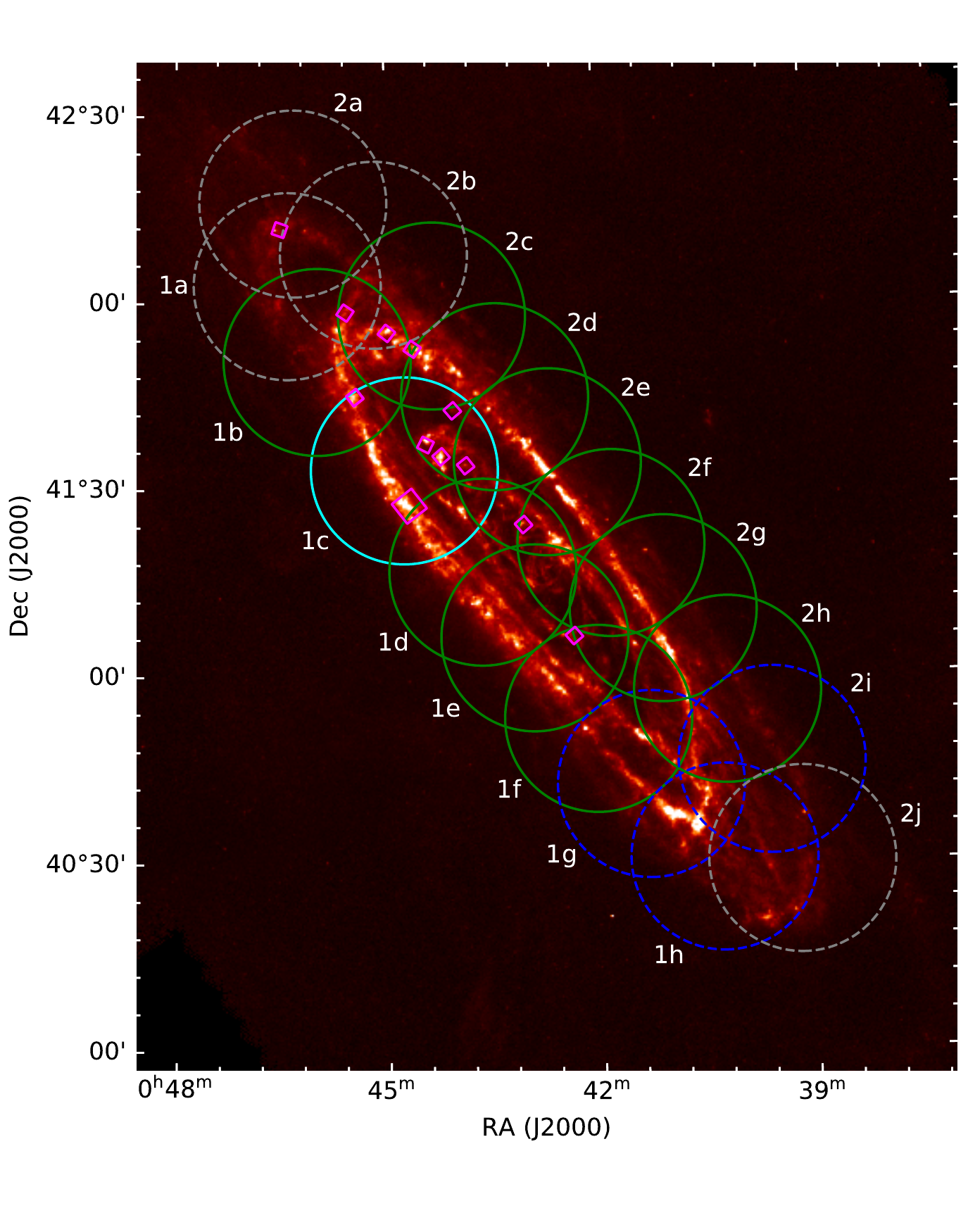}
  \caption{The HASHTAG observing plan superimposed on the {\it Herschel} \SI{250}{\micro\meter} image \citep{smith2012}. Each circle represents one of our 30\arcmin\ {\it Pong} observations, each identified by a white label. The color of the circle represents the observing status of the {\it Pong} when the DR1 products were 
           constructed (November 2020):
           green shows pongs for which all 17 observations have been completed; dashed blue shows 
           roughly half (eight or nine), have been completed; and dashed grey indicates no observation has yet been made (1a has one observation). The cyan circle (1c) is our `pilot' field from 2015 which has 37 observations, which is why there
    	   is less overlap with the other circles. The magenta squares show the regions covered by our CO($J$=3--2) observations.}
  \label{fig:surveyStrategy}
\end{figure}

We continued to use the 30-arcmin {\it Pongs}, choosing the center of each field so that the field contained a similar 
amount of blank sky as the pilot field. This decision required us to have two rows of {\it Pongs} along the major axis. To achieve fairly uniform sensitivity we chose the positions of the centers so the circles
overlapped, as is shown in Figure~\ref{fig:surveyStrategy}. At each position
we made 17 repeat observations of 43 minutes each. Since each position in the galaxy will be covered by at least two {\it Pongs}, every point in the
galaxy will be observed, when the survey is complete, at least 34 times, with
a total integration times of 24.4 hours per 30 arcmin diameter region, the same as the pilot survey. 
We also include data from two significantly shallower 
projects (M12BU26 \& M13BU18) who also observed the entirety of M\,31.

There are two advantages of the design of the fields shown in Figure~\ref{fig:surveyStrategy}. There is a very large amount of
overlap in the observations, especially as the area covered by
a {\it Pong} is somewhat larger than the nominal 30-arcmin circle. This overlap
and redundancy in the data is a considerable help in the data reduction, in particular for
distinguishing real emission from atmospheric emission. The second advantage
is that the two rows of {\it Pongs} overlap, so the sensitivity of the survey will
be much greater along the major axis of the galaxy, which helpfully covers the
central regions of Andromeda where the dust emission
is much weaker than in the star-forming ring.

Inspired by the discovery of luminous transients in the mid-infrared \citep{kasliwal2017}, one of our science goals is to determine whether there
are any luminous transient sources in the submillimeter waveband. We have therefore
split our 17 observations in each field into two sets of 9 and 8 observations,
with the aim of eventually producing two images of Andromeda separated in time, so
we can search for transient phenomena. We have tried to prioritize our
observations to ensure that there is at least a six-month gap between the two sets for
each field, but due to the vagaries of the weather, and the flexible observing queue at the JCMT, 
it is impossible to do this perfectly. However, we  do achieve some time cadence in the observations, which
we will investigate in future works.


\section{Data Reduction: 1. The Method}
\label{sec:dataReduc}

Big sub-millimeter datasets can be challenging to reduce, but HASHTAG is particularly difficult. Large cosmology programs for example 
produce datasets as large as ours but they have the advantage that the data can be reduced piecemeal; the individual datasets are reduced separately and then
the images added together. We were not able to
follow this approach as we want to maximize our sensitivity to extended low signal-to-noise  
emission, which required us to reduce all the data together.

In this section we describe the elements of our method. In Sections~\ref{sec:simulations} and \ref{sec:450sim} we describe the sky simulations we carried out to optimize the method. Unless stated otherwise, the methods we used for the observations and for the data reduction are
the standard ones used for SCUBA-2 \citep{Dempsey2013,Chapin2013}.

\subsection{Initial Processing and Quality Review}
\label{sec:initialProcessing}

As soon as one of our observations was made, we processed it using a quick-look script which produced images at 
the two wavelengths (450 and \SI{850}{\micro\meter}) and saved
the cleaned data from the individual bolometers. We used these images to check for
any severe problems (e.g., array failures or low sensitivity), which for SCUBA-2 data is extremely rare (only 2 out of 235 observations had issues). 
We used the saved bolometer data from this initial processing as the input for the next stage, which resulted in a considerable rationalization of the data, since the initial processing converted the hundreds of raw-data files into just eight (one per array). 

\subsection{An adaptable skyloop: scubaDuperSkyloop}
\label{sec:pipeline}

To reduce SCUBA-2 data, the observatory provides an iterative data-reduction procedure called  
\textsc{makemap}, which at the end of every iteration provides a better estimate of the astronomical signal and the noise until no further improvement is made. Here we provide a brief summary of 
the \textsc{makemap} algorithm but refer the reader to \citet{Chapin2013} for full details. \textsc{makemap} is provided as part of the \textsc{starlink} software package \citep{Currie2014} and throughout this paper we use a recent development build of \textsc{starlink} (version 9072c4434 from April 2020), 
as we required some updates since the last 2018A stable release. 

Initially \textsc{makemap} splits the raw data into individual observations (sub-divided into
chunks if enough RAM is not available, although for HASHTAG `chunking' was not required). There is an initial cleaning step for each observation in which bad bolometers are masked and artifacts (e.g.
glitches) are removed from the timelines. 

\textsc{makemap} then starts the iterative process. At the beginning of each iteration, the common-mode signal (common to all bolometers) is removed. The data
is then corrected for atmospheric extinction, and a high-pass filter is applied to the
timeline data to remove any residual slowly varying signal. Since in the SCUBA-2 arrays are moving across the sky, the removal of a slowly varying signal is equivalent to removing emission on a large angular scale.
An image is then made from the data. Any real astronomical
signal is then identified in the image, and this astronomical signal is then removed from the
timelines (an optional signal-to-noise cut or mask can be applied). The process is then repeated on the new timeline data, with
the astronomical signal being updated in each iteration. 
The process stops when the pixel variations in the map at the end of each iteration fall below a set threshold (i.e., when
the map has `converged'). If we had used the standard implementation of \textsc{makemap}, our final image
of Andromeda would have been a mosaic of the images made by \textsc{makemap} from each
individual observation.

However, a weakness of \textsc{makemap} is that a single $\sim$43 minute \textit{Pong} observation does not have the sensitivity to detect the low-surface-brightness emission
in Andromeda's disk. Recognising this limitation, the \textsc{starlink} team created a script called 
\textsc{skyloop}\footnote{\url{https://starlink.eao.hawaii.edu/docs/sun258.htx/sun258ss72.html}}, 
which runs \textsc{makemap} on all the observations, one iteration at a time, combining the individual images at the end of each iteration
to produce the best estimate of the astronomical signal, thus maximizing the signal-to-noise ratio in the faint extended structure. However, the volume of our
data is so large that \textsc{skyloop} would take too long to run.
We built on our previous work \citep{Smith2019}
to develop a new version of \textsc{skyloop} that we nick-named `scubaDuperSkyloop'. 
The main difference from the old script is that while \textsc{skyloop}
 is given all the observations and internally processes them individually,
 mosaiking the images at the end of every iteration, our modified script calls \textsc{makemap} separately for each iteration and each observation. 
 We found that the new script was more stable on our system (possibly due to more regular memory clearance),however, the main advantage of our approach is that each observation can be processed in parallel, or even on
separate machines. Although \textsc{makemap} can process individual observations using multiple-threads, there are diminishing improvements
in processing time. 

To make the final images shown in this paper, we reduced all the data using a new processing machine with 768\,GB of RAM and 32 
cores/64 threads, which allowed us to run five invocations of \textsc{makemap} in parallel. The main factor limiting the speed of `scubaDuperSkyloop' was the speed of our hard-disk drives. We increased the speed of the simulations described
in section~\ref{sec:simulations} by running the script in parallel on two 
separate smaller machines (with common storage access).

\subsection{Restoring the Large-Scale Structure}
\label{sec:feather}

Ground-based sub-mm surveys of sources with extended emission (greater than a few arcminutes) face the challenge of slow variations in both the atmospheric emission and within the camera which
need to be removed from the data. This is overcome using harsh high-pass filtering,
which also removes real astronomical signal on large angular scales. However, there are sub-millimeter observations of many objects with the space telescopes 
\textit{Planck} and \textit{Herschel}. The images made with these telescopes
do preserve the large-scale structure; however, because of the small sizes of the
telescopes' mirrors they do not have as good resolution as images made with
telescopes on the ground. In principle, we can now produce high-fidelity
images of submillimeter sources with large angular sizes by combining
observations with telescopes on the ground, which provide the small-scale structure, with observations made with telescopes in space, which provide the missing large-scale structure. This technique has often been used in radio astronomy to combine
single-dish and interferometer measurements, as the latter sparsely samples the UV-plane the case here is potentially simpler. 
 
 We have produced the high-fidelity image of Andromeda at \SI{850}{\micro\meter} by combining
 the SCUBA-2 image at  \SI{850}{\micro\meter} and the 353-GHz {\it Planck}, which are
 at virtually the same wavelength. At \SI{450}{\micro\meter}, we have
 combined the SCUBA-2 image at this wavelength with the {\it Herschel} image
 at  \SI{500}{\micro\meter} \citep{smith2012}. To combine the low- and high-resolution images we have written a python module which applies a `feathering' technique \citep{Bajaja1979}. The module performs the following steps:
 
\begin{enumerate}
  \item The low-resolution FITS image is re-projected so that the pixel scale and the celestial coordinates of the pixels are the same as in the high-resolution image. The units of both
  images are converted into Jy\,beam$^{-1}$ using information 
  contained in the header or supplied by the user.
  
  \item We apply color-corrections to the high and low resolution images to correct for the effect of different instrumental filters, 
  calibration schemes (e.g., different reference spectra), and differing central frequencies. To apply this correction the user can specify
  a fixed dust temperature and $\beta$, provide maps of the dust parameters, or a PPMAP cube with the surface-density of dust for a grid of dust temperatures and $\beta$'s. A pre-computed grid specific to each far-infrared/sub-millimeter instrument is used to perform the corrections, and is provided with the task.

  \item The median value in each image is subtracted from the image and `NaN' pixels are replaced with zeros to avoid artifacts.
  
  \item Both images are Fourier transformed and shifted so a spatial frequency of zero is assigned to the center. The values in the Fourier Transform (FT) of the low-resolution
  image are scaled by the ratio of the beam areas of the high-resolution and low-resolution images.
  
  \item A filter is then created to weigh the FT images by
  the selected amount when they are combined. The standard filter in the module,
  which we used for HASHTAG, is a Gaussian filter in Fourier space. We chose the
  value of the filter's standard deviation using the simulations described
  in the next section.   
  The FT of the high-resolution image is
  multiplied by one minus the Gaussian filter. 
  The FT of the low-resolution image by default is not  
  weighted, as the image is effectively already weighted by the point spread function of the image; this is the same method employed by \textsc{casa} \citep{casa}. 
  In our simulations we also try the alternative where the low-resolution FT image 
  is weighted by the Gaussian filter, which is applicable when the filter is on significantly larger spatial scales than the 
  resolution of the images. For a full discussion of combining images in the Fourier plane see \citet{Stanimirovic2002}.
  The weighted FT images are then added together. There is also an option in the module to use either a Butterworth \citep{Csengeri2016} or Sigmoid filter, in which case both FTs are multiplied
  by the filter. We tried these filters for HASHTAG but found they did not
  produce appreciable 
  benefit over the Gaussian filter.
  
  \item The combined FT image is then inverse Fourier transformed (with appropriate inverse shifts applied). The 
  NaN pixels are restored, the median background from the original
  low-resolution image is added back to the new image, and the keywords in the image header are updated\footnote{There is an option in the module to add back a 
  background by calculating the offset between the original low-resolution image and the new image smoothed to the same resolution.}.
  
  \item As many SED fitting procedures apply a color-correction step in their processing, we remove the color corrections performed in step 2, so the flux densities in the final images are based on the same assumptions as regular
  SCUBA-2 images.
\end{enumerate}

One key parameter in the `feathering' algorithm is the scale of the Gaussian filter. As the feathering step is a relatively quick process (compared to the
SCUBA-2 pipeline) we keep this as a free parameter which we optimize in Section~\ref{sec:simulations} and \ref{sec:450sim}.
For the color corrections  we assume the spectral energy distribution (SED) in each pixel estimated by \citet{Whitworth2019}, 
who applied the PPMAP algorithm to the {\it Herschel} dataset to generate SEDs with the angular resolution of the
highest resolution {\it Herschel} image. 
For the {\it Herschel} and {\it Planck} images, we calculated the color corrections using these SEDs and the filter curves available on the observatories' websites, after removing the standard SED used to estimate the {\it Planck} and {\it Herschel} flux densities ($F_{\nu} \propto \nu^{-1}$).
SCUBA-2 flux densities, on the other hand, are calibrated relative to Mars and Uranus, which means that they are based on a very different assumption about SEDs \citep[roughly $F_{\nu} \propto \nu^{1.7}$,][]{Lellouch2008,Orton2014}. We calculated the color corrections for the SCUBA-2 images after removing this assumption and then using the PPMAP SED in
each pixel. For SCUBA-2, an additional complication is that the effective filter
function is the product of the actual filter function and the atmospheric
transmission. We calculated the effective filter function from the
filter function available on the observatory's website and a model for the
transmission of the atmosphere\footnote{Atmospheric model from from CSO \citep{Pardo2001}, \url{http://www.submm.caltech.edu/cso/weather/atplot.shtml}.} with $\tau_{\rm 225\,GHz}=0.065$ to match the weather for our survey. The color corrections are small ($\leq$3\%) for the two SCUBA-2 filters and for the {\it Herschel} \SI{500}{\micro\meter} filter, apart from the large correction
needed to change the flux at \SI{500}{\micro\meter} to one at \SI{450}{\micro\meter}, which ranges
from  a factor of $\sim$1.5 in the center to 
$\sim$1.34 in the ring. The color correction for the {\it Planck} filter
is $\simeq$10\%.

\begin{figure}
  \centering
  \includegraphics[trim=18mm 3mm 3mm 3mm, clip=True, width=0.49\textwidth]{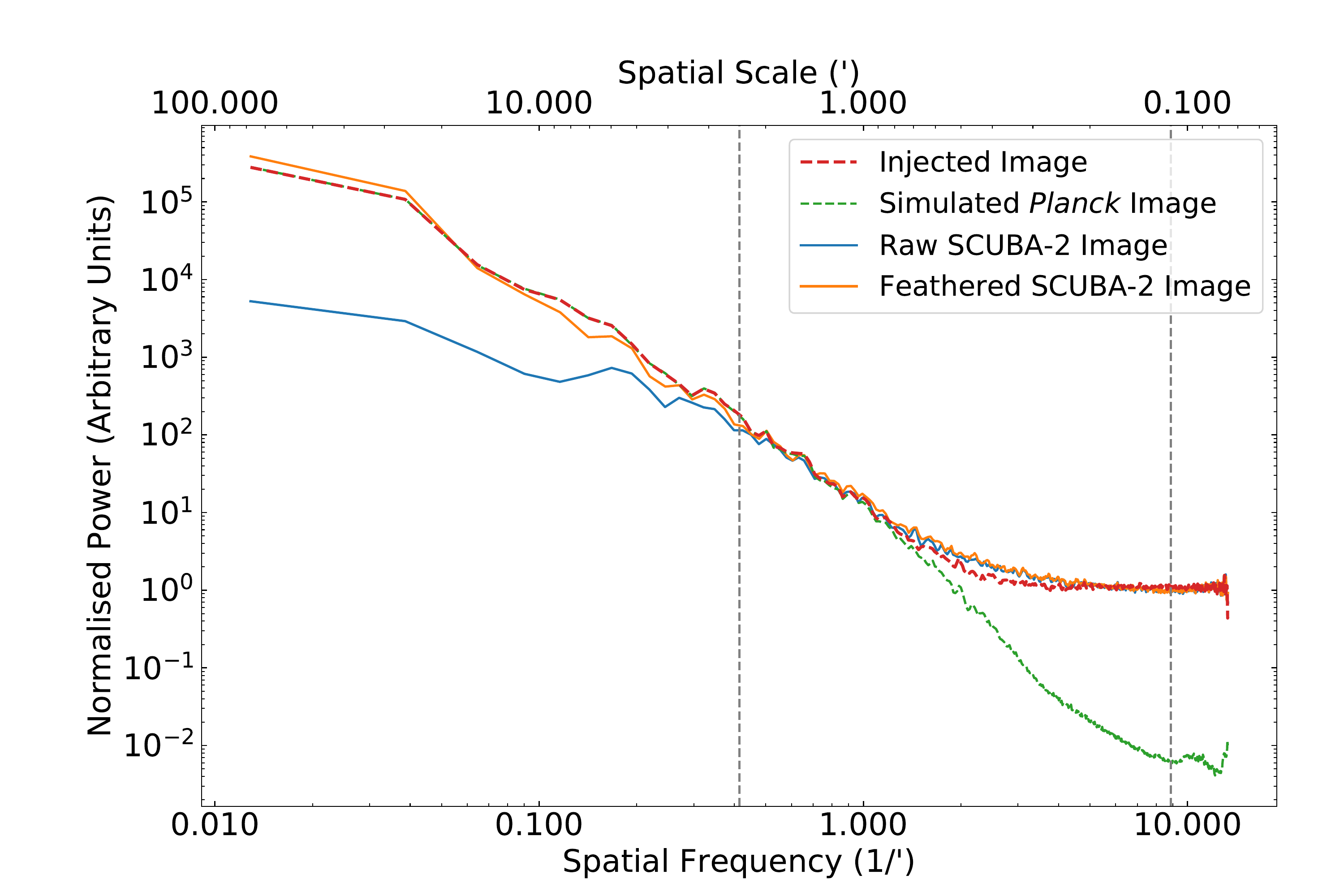}
  \caption{A demonstration of the ability of our feathering technique to recover the
  structure of the sky on all spatial scales, using the final simulation
  in Section~\ref{sec:simulations}. The red line shows the power spectrum of a simulated `true' image of the sky at \SI{850}{\micro\meter} (with artificial noise added to match the SCUBA-2 image). The green line shows the power spectrum of what {\it Planck} should see, obtained by convolving the true
  image to the {\it Planck} resolution. The blue line shows what SCUBA-2
  should see, obtained by passing the true image through the SCUBA-2 data-reduction
  pipeline. As expected, the {\it Planck} spectrum is missing the high-\textit{k} Fourier
  components and the SCUBA-2 spectrum the low-\textit{k} Fourier components.
 The orange line shows the power spectrum of our
reconstruction of the sky by applying our feathering technique to the artificial
{\it Planck} and SCUBA-2 images. For reference the location of the half-width half-maximum (HWHM) for \textit{SCUBA-2} and 
\textit{Planck} is shown by the grey vertical lines. Note in all cases we have 
           restricted the images to the central 30\arcmin\ region. 
           }
  \label{fig:powerSpectrum}
\end{figure}

Figure~\ref{fig:powerSpectrum} shows the results of the feathering technique when applied to SCUBA-2 simulation (see Section~\ref{sec:simulations}),
illustrating how it is effective at restoring the structure on all spatial scales.
The red line shows the power spectrum of a simulated `true' image of the sky at  \SI{850}{\micro\meter}. The green and blue lines show models of what {\it Planck} and SCUBA-2
would see, respectively, in one case missing the high-\textit{k} and in the other case
the low-\textit{k} Fourier components. The orange line shows the power spectrum of our
reconstruction of the sky by applying our feathering technique to the artificial
{\it Planck} and SCUBA-2 images.

We performed a sanity-check of the method by applying it to a {\it Herschel}  \SI{250}{\micro\meter} image
of the Milky Way \citep{molinari2016}, in which the large-scale emission is detected
with high signal-to-noise. We created an artificial image at the {\it Planck} resolution by smoothing
the {\it Herschel} image, and we created
a rough simulation of what a camera like SCUBA-2 would see by using the
\textsc{Nebuliser} algorithm developed by the Cambridge Astronomical Survey Unit\footnote{http://casu.ast.cam.ac.uk/surveys-projects/software-release/background-filtering} to remove the
structure on large scales. When we produced our combined image by
applying our feathering technique to the two artificial images, we found a
good agreement with the original image (to within a few percent), although in the brightest regions there
were differences up to the 10\% level. 

There are also some parameters to tune in \textsc{makemap}, in particular
the scale of the high-pass filter used to remove the signal from the atmosphere
and the camera itself. To produce a reliable map of Andromeda it is crucial to
get these values right. This is a particular challenge at the
longer of the two SCUBA-2 wavelengths because of the need to ensure that emission on the SCUBA-2 images
is preserved on all angular scales up to the angular resolution of {\it Planck} ($\simeq$5 arcmin, FWHM). In the next section, we describe the simulations of the sky that we used to determine the best values of the parameters.

\section{The Simulations at \SI{850}{\micro\meter}}
\label{sec:simulations}

We carried out simulations of the sky to optimize the data-reduction procedure
described in the previous section. The SCUBA-2 pipeline has many parameters which can 
be tweaked to optimize the reduction process depending on the angular extent of the source, observing
strategy and the atmospheric conditions. Two of the big unknowns are the scale of the
Gaussian filter used in combining the low- and high-resolution data and the
angular scale of the high-pass filter that should be used in the SCUBA-2 data reduction to remove
the noise on large angular scales caused by the atmosphere and the camera. Too harsh a filter would remove the
noise but also remove too much astronomical signal, too weak a filter would leave the
astronomical signal alone but not remove the noise. In this section we create a `simulation' to test the effects of
the various pipeline parameters so we can obtain the most accurate map of M\,31's sub-millimeter emission. In this process
we have restricted ourselves to the SCUBA-2 pipeline, rather than attempt alternative methods 
\citep[e.g., \textsc{Scanamorphos},][]{Roussel2013} or complex atmospheric modelling.

An outline of our method is as follows. (1) We used real SCUBA-2 data from a cosmology Large Programme with similar noise properties to our own dataset. (2) We then created a `true' image of Andromeda from a {\it Herschel} image and inserted this into the timelines for the cosmology programme. (3) We convolved the true image to produce an artificial {\it Planck} image. (4) We ran the SCUBA-2 dataset (cosmology timelines injected with our model galaxy)
through the SCUBA-2 data-reduction pipeline (Section~\ref{sec:pipeline}). (5) We combined 
the reduced SCUBA-2 image and the {\it Planck} image to try to recover the
original image using the method of Section~\ref{sec:feather}.
(6) We measured the statistical differences between our recovered image of Andromeda
and the original true image. We ran thousands of simulations, trying different variants of
the SCUBA-2 data-reduction pipeline, in particular trying different values of the
scale of the high-pass filter, and trying a range of values for the scale of the Gaussian
filter used to combine the low-resolution and high-resolution data (Section~\ref{sec:feather}).
In this section we describe the simulations at \SI{850}{\micro\meter}, which were more critical
because {\it Planck} at  \SI{850}{\micro\meter} has a much lower resolution than
{\it Herschel} at  \SI{500}{\micro\meter}. We describe what we did for the  \SI{450}{\micro\meter} data
in Section~\ref{sec:450sim}.

\subsection{Simulation Setup}

We chose to carry out a simulation of a single 30\arcmin\ {\it Pong} observation
with a similar sensitivity to our pilot field. While ideally we would have
simulated observations of the whole galaxy, the processing time would have been too
long and there was no suitable SCUBA-2 data we could use for a simulation of
the entire galaxy at our depth. For our simulation we used data from the SCUBA-2 Cosmology Legacy Survey \citep{geach2017}. We chose to use the data from the survey of the Lockman
Hole because it was carried out in similar weather conditions and
consisted of 35 30\arcmin\ {\it Pongs}, the same as we used in our observations of our pilot field (Section~\ref{sec:surveyStragey}), reaching a similar sensitivity.

We made our `true' image out of the {\it Herschel}  \SI{250}{\micro\meter} image, which has
a resolution (18 arcsec, FWHM) which is not very different from that of
SCUBA-2 at  \SI{850}{\micro\meter} (13.5 arcsec, FWHM). We first reprojected the {\it Herschel}
image onto a 4\arcsec\ pixel grid
and then multiplied the intensity values in each pixel by the
ratio of the global \SI{250}{\micro\meter} and  \SI{850}{\micro\meter} fluxes \citep{planck2015}. The Lockman-Hole
data are actually slightly less sensitive than the data for our pilot field
(3.4 versus 3.0\,mJy\,beam$^{-1}$), so to make the signal-to-noise ratio of our artificial
image the same we multiplied the intensity values by the ratio of the noises.
We also applied color corrections. We then converted the
intensity values into instrumental units of pico watts (pW) and ran
\textsc{makemap} with the options ``fakemap'' and ``exportclean'' set so that our artificial \SI{850}{\micro\meter} image was added to the SCUBA-2 timelines for the Lockman
Hole survey. We used the data files exported by this process, which also did
the initial cleaning of the timelines, to carry out our simulations.
We produced our artificial {\it Planck} image by convolving our
input image to the {\it Planck} resolution.

We performed five different sets of simulations, to optimize different aspects of the method. While in an ideal world every possible combination of parameters
in the SCUBA-2 data-reduction pipeline would be tested, each individual simulation required $\sim$8--10\,hours to run.
Therefore, we varied one parameter at a time. However, our reconstruction
process in which we combined the processed SCUBA-2 image with our
artificial {\it Planck} image was quite quick to run, so at the end of every
run of the pipeline, we applied our reconstruction technique many times, each
time with a different value for the scale of the Gaussian filter, with scales
ranging from 160 to 840\arcsec.

We measured the success of a simulation by the differences between
our final image and the original true image. We assessed the significance
of the differences using the noise image produced by the pipeline, but since changes
to the pipeline also change the noise in the final image we also used a reference
noise image (from a run using typical values of all the pipeline parameters).
For each final SCUBA-2 image, we created several `difference maps' of
the difference between the image and the true image: (1) a basic residual image, i.e., the final image minus the `true' image; (2) the residual image divided by the noise image; (3) the residual image divided by the reference noise image; (4) the residual
image divided by the true image. 

To assess the agreement, we measured statistics in two different regions of these
difference maps:
(1) the full-depth region of the {\it Pong}; (2) the full-depth region of
the {\it Pong} but only for pixels where the flux in the pixel
in the {\it Herschel}  \SI{500}{\micro\meter} image is above a critical value
(\SI{500}{\micro\meter} is used as its closest in wavelengths to both SCUBA-2 bands).
The point of the second region was to stop the more numerous pixels outside the 
galaxy with little emission biasing the results, as a method producing a flat map
(e.g., harsh Fourier filtering) would be preferred.
We inspected all of these methods
of assessing the agreement at one time or another.
The statistic that we found most useful was the mean of the absolute difference
between the final image and the true image divided by the reference noise image,
for pixels in the deep region above the \SI{500}{\micro\meter} threshold. This statistic
using a reference noise map has the advantage that changes in the noise map do not bias the estimate
of how well we recover the galaxy, while still accounting for variations in the sensitivity across the map.

\subsection{Filter-scale and PCA-components}
\label{sec:PCAsims}

The most important parameter in the SCUBA-2 data-reduction pipeline is the scale of the high-pass filter used to remove residual emission from the atmosphere or
the instrument. We started our simulations with the expectation that we would need to set
the scale to roughly the angular resolution of {\it Planck}. In our early results we found that with a filter scale of 340\arcsec\ we were able to reproduce the true image well. 

In April 2019, however, the observatory released a new mode for the
SCUBA-2 pipeline in which principal component analysis (PCA) is
used to remove residual atmospheric and instrumental noise. The advantage of PCA is that it makes it possible to increase the
angular scale of the high-pass filter, reducing the attenuation of the
emission on the SCUBA-2 images on large angular scales. Of course, if one
allows PCA to remove too many components, it is also possible to remove
real astronomical emission from the image. In the SCUBA-2 pipeline the default number of PCA components
to remove is 20 components per sub-array, but after inspecting the final
images made with this setting we decided it led to the removal of
some real emission.
We therefore decided to run a suite of simulations in which we varied both the
scale of the high-pass filter and the number of PCA components.

\begin{figure}
  \centering
  \includegraphics[trim=0mm 0mm 0mm 0mm, clip=True, width=0.49\textwidth]{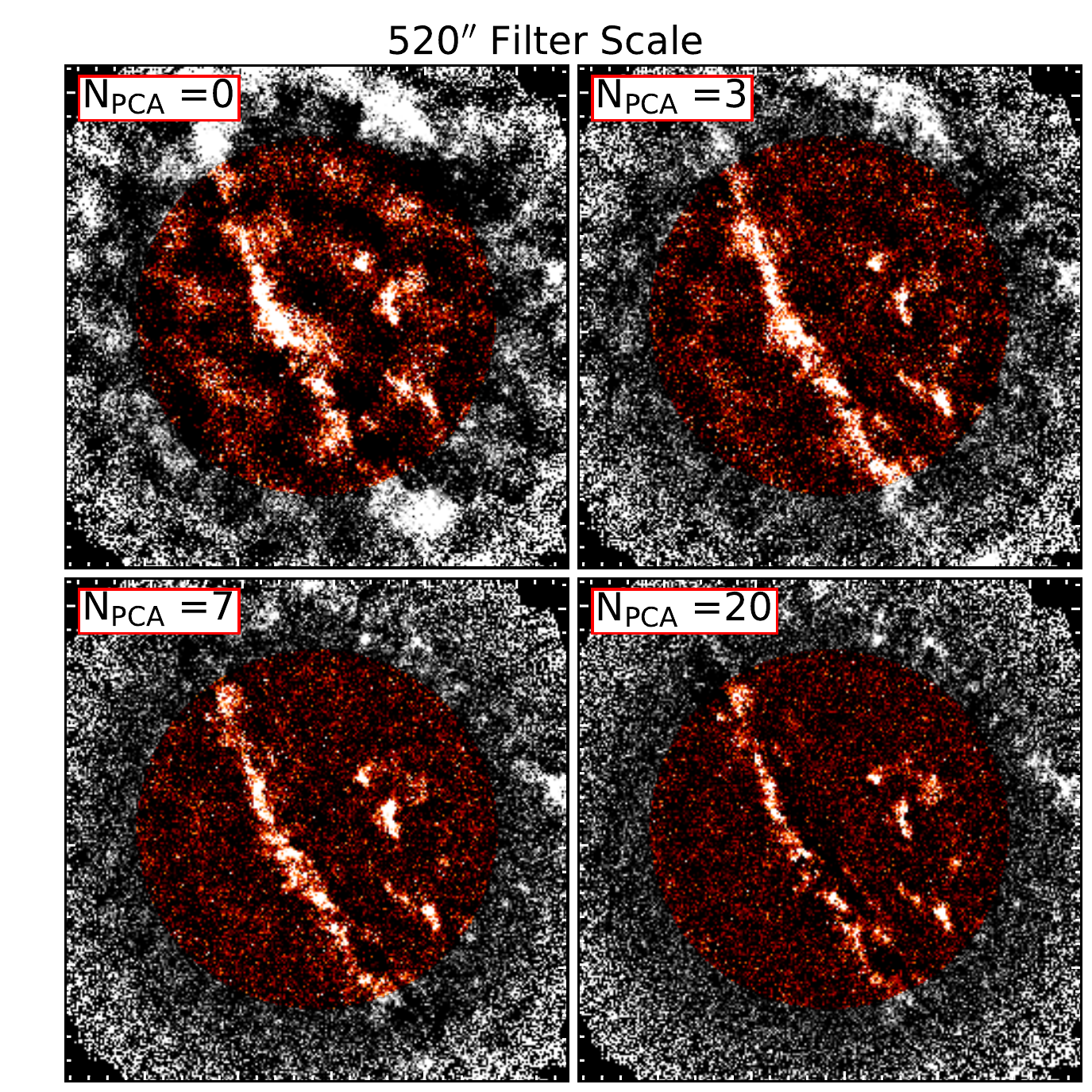}
  \caption{Reconstructed SCUBA-2 images made with a high-pass filter in the SCUBA-2
  pipeline of 520 arcsec but with different numbers of PCA components removed
  (0, 3, 7 and 20). The red area is the circular region with a diameter
  of 30\arcmin\ in which the {\it Pong} reaches full sensitivity. The color
  scale has been chosen to enhance faint features. The removal of
  too few PCA components leads to large-scale artifacts in the image (top left panel); too many
  PCA components removes real astronomical signal, leading to the
  negative regions close to the bright structure (bottom right panel).
 }
  \label{fig:PCAcomp}
\end{figure}

In our initial simulations without PCA we had found an optimum filter scale
of 340 arcsec. We realized that with PCA we should be able to increase this scale.
We therefore ran simulations with filter scales between 340\arcsec\ and 560\arcsec\
and the number of PCA components between 0 and 20, running 91 simulations
to cover this 2D parameter space (20 PCA components is the default in the new pipeline mode).

Figure~\ref{fig:PCAcomp} shows the effect of changing the number of PCA components 
while keeping a constant filter scale of 520 arcsec. The top-left hand panel shows
what happens if no PCA components are removed. The filter may not remove
much large-scale emission from the galaxy but much of the emission visible
in the picture is clearly spurious. Increasing the number of PCA components leads to better removal of these artifacts, but the removal of too many PCA components
leads to the removal of real astronomical signal, producing the negative regions around
the image in the bottom-right panel, which has had 20 PCA components removed.

\begin{figure}
  \centering
  \includegraphics[trim=5mm 6mm 8mm 8mm, clip=True, width=0.49\textwidth]{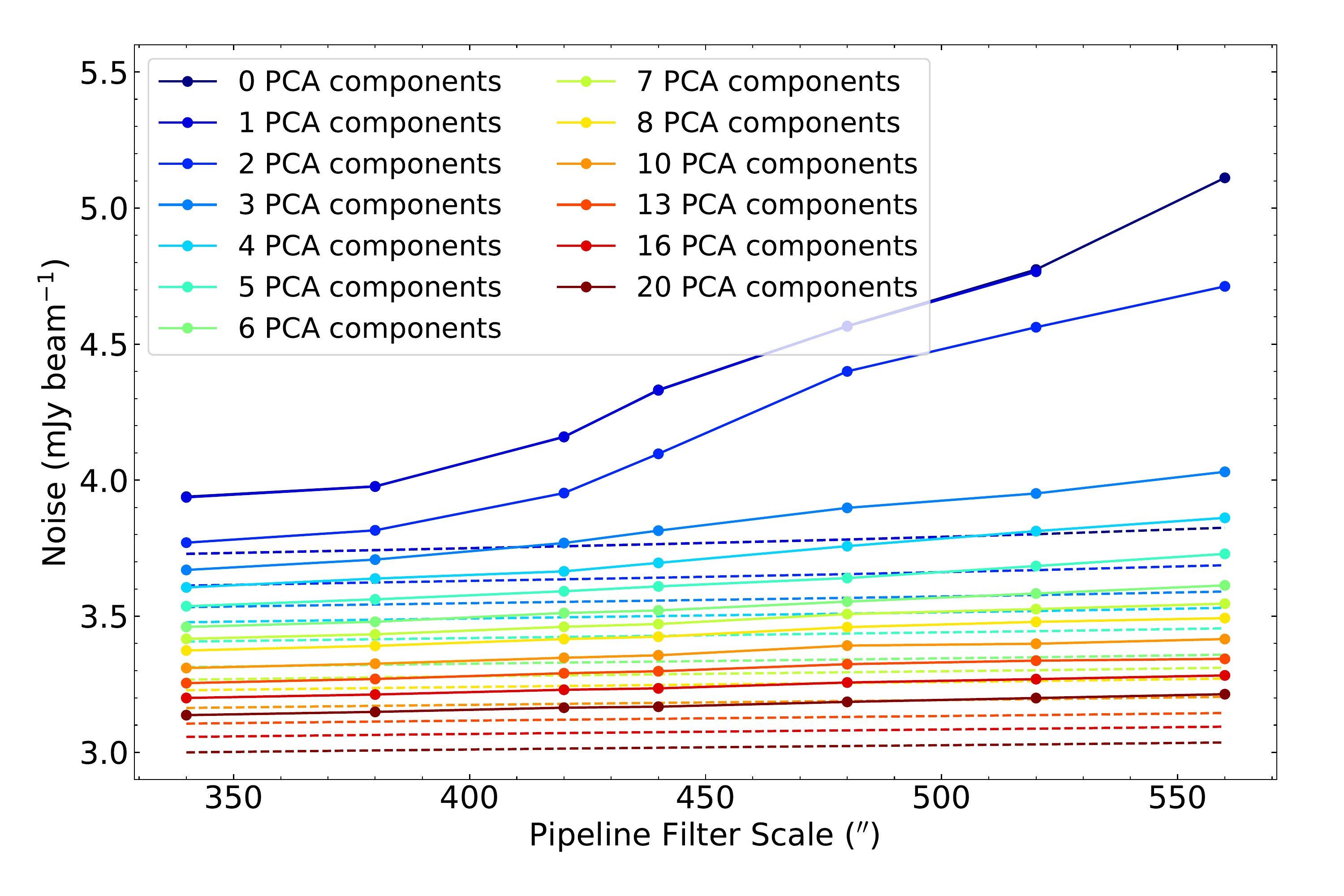}
  \caption{The relationship between the average noise in an image as a function of the
the scale of the 
high-pass filter and the number of PCA components that have been removed. The solid lines show the results when the noise
is measured from empty areas of the image and the dashed lines show when
it is measured from the noise image produced by the SCUBA-2 data-reduction
pipeline. The difference between the two sets of lines shows that the
noise values from the variance map produced by the pipeline are unreliable because the pipeline mistakenly
identifies large-scale noise as real astronomical signal and therefore underestimates the noise. Even when these artifacts are removed
with a harsh high-pass filter or by removing a large number of PCA components,
there is still a small difference of $\sim$0.2\,mJy\,beam$^{-1}$, which is
possibly caused by faint sources in the apparently empty regions of the image.
}
  \label{fig:noisePlot}
\end{figure}

An important point to note is although the eye tells us that
the patches in the top left panel in Figure~\ref{fig:PCAcomp} are artifacts, the SCUBA-2
data-reduction pipeline treats these as real emission, which means that
the values in the noise image produced by the pipeline are too low.
Figure~\ref{fig:noisePlot} shows how the average noise on an image depends on the
number of PCA components and on the scale of the 
high-pass filter. The solid lines show the results if the noise is measured
from empty areas of the final image; the dashed lines show the result
if the noise is measured from the noise image produced by the SCUBA-2 data-reduction
pipeline. The pipeline estimates are much lower, showing the effect of the pipeline
treating the artifacts as real astronomical signal (Users
of the SCUBA-2 pipeline beware!). The more reliable noise values, measured from the images themselves, show that the noise in an image can be reduced either by
decreasing the scale of the high-pass filter or by increasing the number of PCA
components. In either case, of course, one also runs the risk of removing real
astronomical signal.

\begin{figure}
  \centering
  \includegraphics[trim=5mm 6mm 8mm 8mm, clip=True, width=0.49\textwidth]{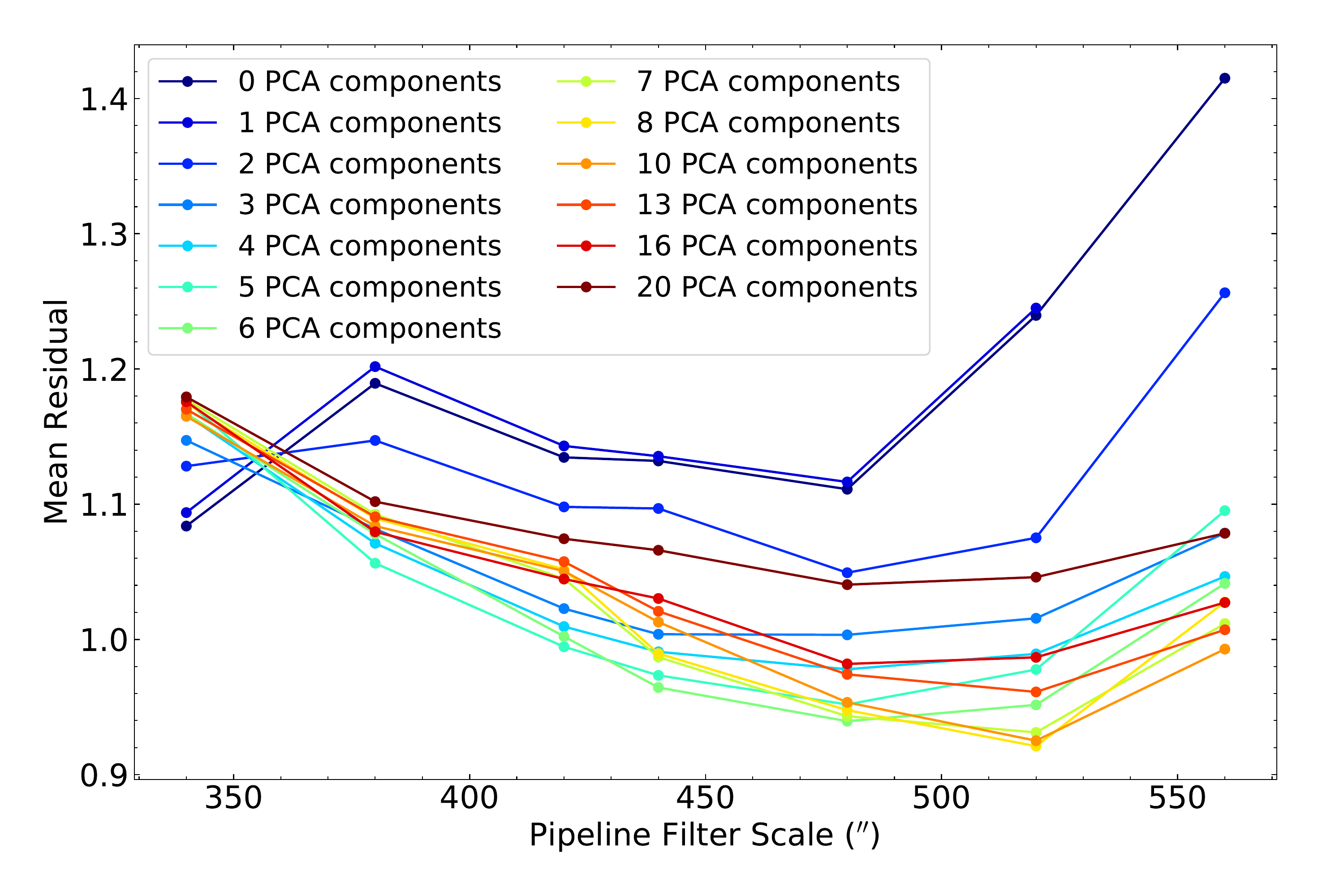}
  \caption{The mean absolute residual in the difference map versus the
  angular scale of the high-pass filter. Each line shows the results when a different
  number of PCA components are removed. The difference map is the residual image
  (true -- recovered) divided by the reference noise image and pixels
  have only been included if they lie in the central region (30\arcmin\ diameter)
  of the {\it Pong} and if the \SI{500}{\micro\meter} flux in the {\it Herschel} image
  of Andromeda \citep{smith2012} is greater than 200\,mJy\,beam$^{-1}$.
  The figure shows the best combination (smallest difference between input
  and output images) is a filter scale of $\sim$480--520\arcsec\ and
  5--10 PCA components.
  }
  \label{fig:meanResidual}
\end{figure}

We chose the best combination of filter scale and number of PCA components
based on a combination of visual inspection of the residual maps and the
statistical estimates of the difference between the recovered image and the
true image. Figure~\ref{fig:meanResidual} shows the mean absolute residual
for the difference map made by dividing the residual image by the reference noise image (number three
in the list above). The statistic has been calculated for the pixels
in the full-depth region and which are above the \SI{500}{\micro\meter} threshold value (see above).

The figure shows that there is an advantage in increasing the scale
of the high-pass filter from the 340 arcsec that we had originally
considered (see above) if removal of PCA components is included in
the analysis. The best agreement between the recovered and true image
is obtained for a filter scale of $\simeq$520 arcsec. The figure also
shows that there is an optimum number of PCA components of $\simeq$8 ---
removing more increases the difference between the recovered and true
images. Based on these results, we visually inspected the
recovered images for a filter scale of 480\arcsec\ and 5 or 6 PCA components and 
for a 
filter scale of 520\arcsec\ with 7 or 8 PCA components, concluding
that the best results came with a filter scale of 520\arcsec\ with 7 PCA components. 
However, in the next stage of the simulations we also tested
the method with a filter scale of 480\arcsec\ and 5 PCA components, in order to
check whether the optimum values shifted if other parameters in the method were varied.

In the simulations, we also checked the number of iterations in the pipeline
required for each set of parameters, since the computer processing time
is directly proportional to the number of iterations. We found, as expected,
that there is a processing cost to setting a larger filter scale (from 8 to 19 iterations)
but including PCA analysis can reduce this.

\subsection{The Mask}
\label{sec:maskSim}

An important element in the SCUBA-2 data-reduction pipeline is a mask
provided by the user as their best guess of the area in which real
astronomical signal will be found. This greatly helps the convergence
of the iterative procedure as it makes it easier for the algorithm
to distinguish between real astronomical emission and extended structures
that are actually the result of atmospheric emission or noise in the
camera (see Figures~\ref{fig:PCAcomp} \& \ref{fig:noisePlot}).
This does not mean that a pixel in the final image that is not within the mask will
necessarily contain no astronomical signal. The software only subtracts astronomical
signal from samples in the timelines that will contribute to
image pixels within the mask, but once
all data-reduction stages are completed (subtraction of PCA
components etc.) even pixels outside the mask, which are the averages
of many samples in the timeline, may still contain astronomical signal.

The mask we used was created from the {\it Herschel} \SI{500}{\micro\meter} image of Andromeda
\citep{smith2012}, the {\it Herschel} image closest in wavelength to
our  \SI{850}{\micro\meter} image. We defined the mask as all pixels above a
\SI{500}{\micro\meter} flux-density threshold, with this threshold providing another knob we could
twiddle in our analysis. Too high a threshold makes it possible for
the algorithm to treat real astronomical signal as noise and remove
it from the timelines; too low a threshold leads to slower convergence of the algorithm
and higher noise.
In the simulations described in the
previous section, we set the \SI{500}{\micro\meter} flux-density threshold at 200\,mJy\,beam$^{-1}$, which
seemed a reasonable compromise as the mask
then included most of the disk and some inner regions of the galaxy while
excluding some regions between the rings where the emission is faint.

\begin{figure}
  \centering
  \includegraphics[trim=1mm 12mm 1mm 8mm, clip=True, width=0.47\textwidth]{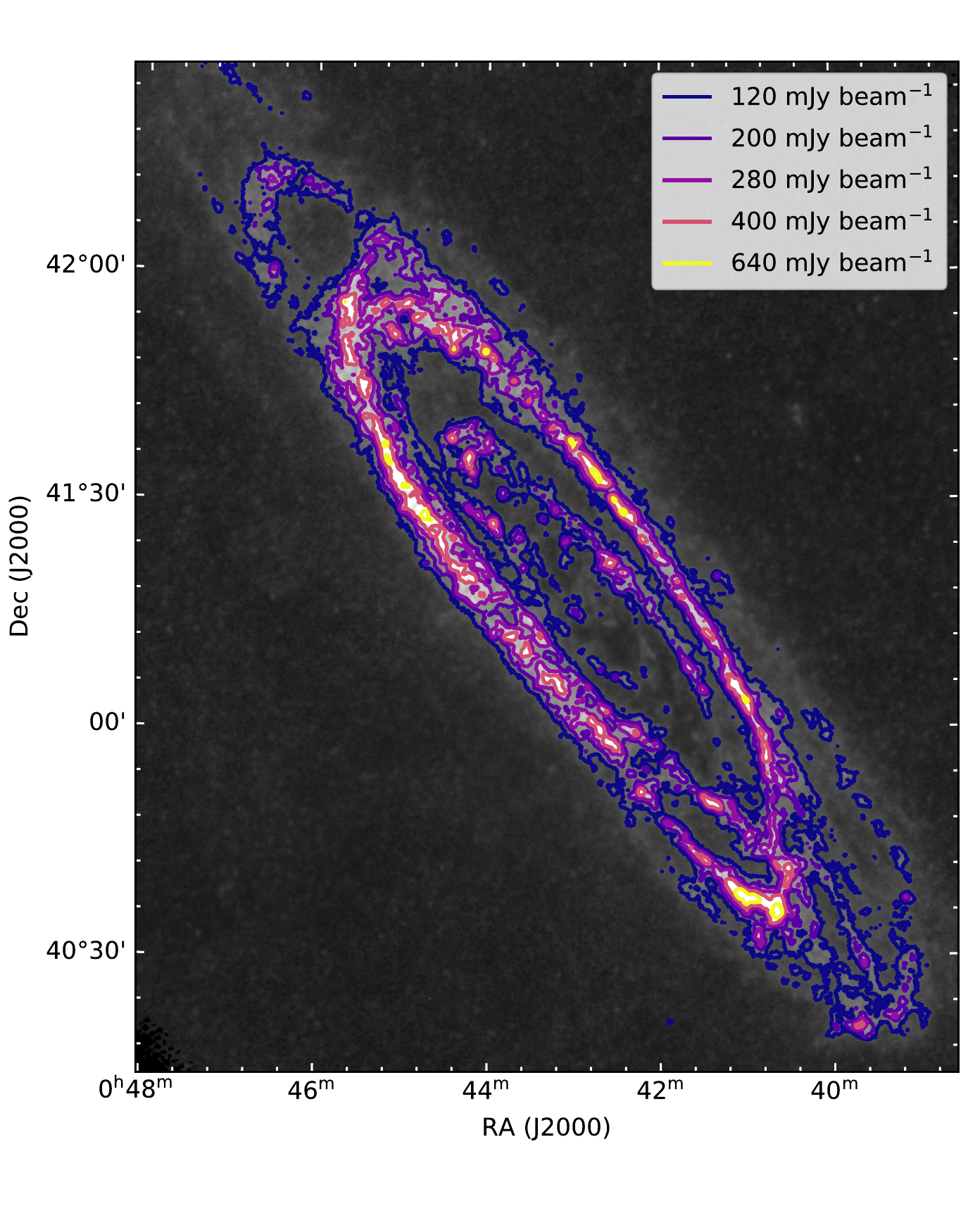}
  \caption{The grey-scale image is the \SI{500}{\micro\meter} {\it Herschel} image 
  which we used to create the mask used in the SCUBA-2 data-reduction pipeline. We
  defined the mask as all pixels with a \SI{500}{\micro\meter} flux density greater
  than a threshold value. In the simulations described in Section~\ref{sec:maskSim}, we
  tested the effect of changing this threshold value. The five contours
  show the masks created from flux thresholds that are representative of the
  ones we used in the simulations.
  }
  \label{fig:maskLevels}
\end{figure}

\begin{figure}
  \centering
  \includegraphics[trim=5mm 8mm 8mm 8mm, clip=True, width=0.49\textwidth]{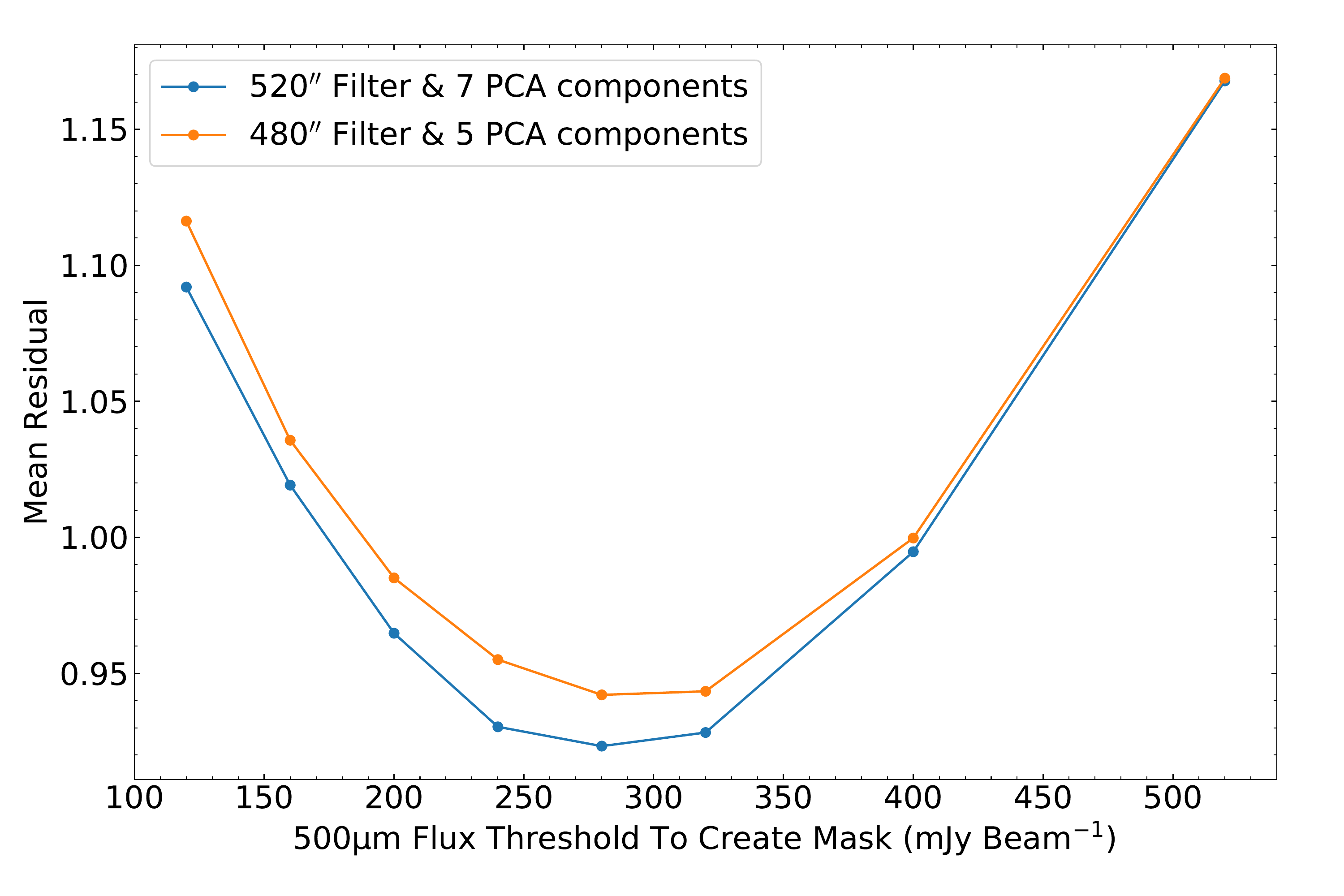}
  \caption{Mean absolute residual in the difference map versus
  the \SI{500}{\micro\meter} flux-threshold used to define the mask used in the
  SCUBA-2 data-reduction pipeline (see the caption of Figure~\ref{fig:meanResidual}
  for details of the difference map and of the region used to measure
  the statistic). The two lines are the results from using the best
  combinations of filter scale and number of PCA components identified during
  the simulations described in
   Section~\ref{sec:PCAsims}.
           The plot shows the best \SI{500}{\micro\meter} flux-threshold for creating
           a mask is 280\,mJy\,beam$^{-1}$ for both combinations of filter
           scale and PCA number.}
  \label{fig:maskResidual}
\end{figure}

Once we had identified the best combinations of filter scale and
number of PCA components (520 arcsec and 7 PCA components or 480
arcsec and 5 PCA components), we used these to test the effect of
changing the flux threshold used to construct the mask. We ran simulations
with values of the \SI{500}{\micro\meter} flux threshold between 120 and 520\,mJy\,beam$^{-1}$.
Figure~\ref{fig:maskLevels} shows masks created 
with a range of flux thresholds that are representative of the ones we
used in the simulations.
Figure~\ref{fig:maskResidual} shows the results of the simulations. The agreement
between the input and output images is clearly best for
a \SI{500}{\micro\meter} flux-threshold of 280\,mJy\,beam$^{-1}$ (this includes $\sim$25\% of the total
flux of M\,31),
which is the threshold we adopted to create the mask for the
real \SI{850}{\micro\meter} observations of Andromeda. We found that changing the
flux threshold and thus the mask had a negligible effect on the
noise of the final image, a very slight increase ($<1$\%) for masks generated
with a \SI{500}{\micro\meter} flux-threshold below 200\,mJy\,beam$^{-1}$.

\subsection{Tolerance Level}

The next parameter we investigated was the map-tolerance parameter which 
the SCUBA-2 data-reduction pipeline uses to decide whether the algorithm has
converged or whether more iterations are required. 
The default tolerance value is 0.05, which means that the iterations stop
when the average change in the flux in a pixel from the last iteration is less than 0.05\,$\sigma$, 
$\sigma$ being the noise in that pixel (calculated from the distribution of instrument samples contributing to that pixel). There have, however, been
some studies  \citep{Mairs2015,Smith2019}
that suggest a lower tolerance value might improve results, something we wanted to
explore in our simulations.
 
Given our huge volume of data, another important consideration was processing
time, which increases if the tolerance value is reduced because a greater number
of iterations are then needed to reach a lower tolerance value (processing
time scales linearly with number of iterations). We therefore 
carried out simulations over a fairly small range of tolerance values: 0.0075--0.05.
In Section~\ref{sec:maskSim} we found 
that we achieved the best results with a mask generated with a \SI{500}{\micro\meter} threshold of
280\,mJy\,beam$^{-1}$. In the simulations described in this section, we also
experimented with masks created with  \SI{500}{\micro\meter} thresholds
of 200 and 240\,mJy\,beam$^{-1}$ to see whether the choice of best mask changed
if we also changed the tolerance parameter. We also tried both winner
and runner-up from the competition between filter-scale/PCA combinations of
Section~\ref{sec:PCAsims} to see if the order might be reversed with a different value
of the tolerance parameter.
 
\begin{figure}
  \centering
  \includegraphics[trim=5mm 8mm 8mm 8mm, clip=True, width=0.49\textwidth]{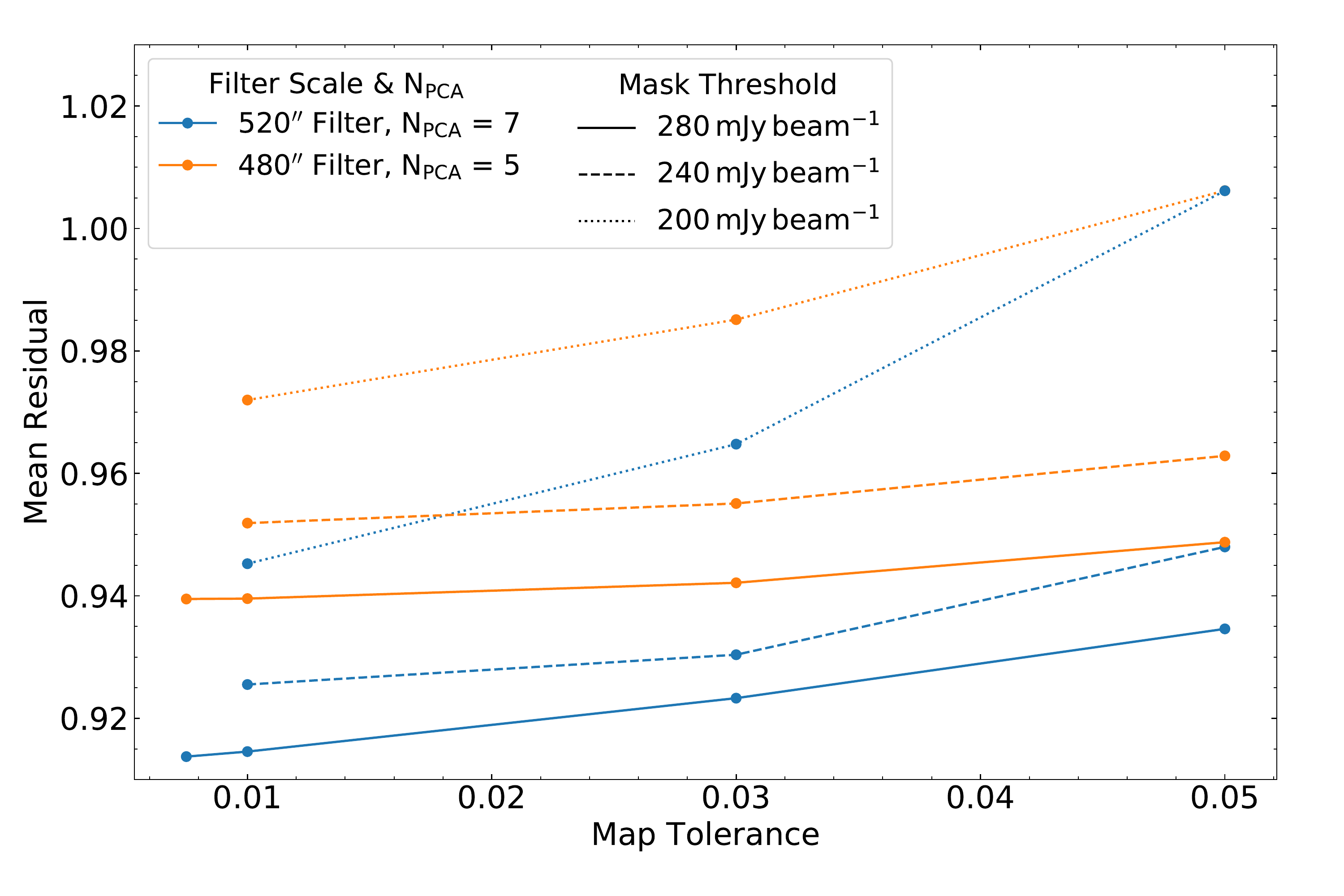}
  \caption{Mean absolute residual in the difference map versus
  the value of the tolerance parameter used in the
  SCUBA-2 data-reduction pipeline (see the caption of Figure~\ref{fig:meanResidual}
  for details of the difference map and of the region used to measure
  the statistic).
  The blue lines 
           show the results from simulations with a filter scale of 520\arcsec\ and 7 PCA components, and the orange lines show the results from simulations with a 
           filter scale of 480\arcsec\ and 5 PCA components. The solid, dashed and dotted lines show the results from simulations with a mask
           made with a \SI{500}{\micro\meter} flux threshold of 280, 240 and 200\,mJy\,beam$^{-1}$, respectively. The plot shows that there is an
           improvement made by reducing the value of the tolerance parameter, but the
           improvement is modest, and there is a trade-off with an increase
           in computer processing time.}
  \label{fig:mapTol}
\end{figure}

Figure~\ref{fig:mapTol} shows that decreasing the value of the tolerance parameter does
improve the agreement between the true and recovered images. They also
show that the best choices for mask and filter-scale/PCA combination
generally remain the best choices at all values of the tolerance
parameter. The improvement with decreasing tolerance parameter is fairly
slow, and there is a high price in increased processing time.
Therefore, as a trade-off, we adopted a tolerance parameter
of 0.03 for the real observations. Even with this tolerance parameter,
reducing the current HASHTAG  \SI{850}{\micro\meter} dataset, which is only 70\% of
the final dataset, required 7.5 days of computer processing time.

\subsection{Feather Scale}

\begin{figure}
  \centering
  \includegraphics[trim=5mm 8mm 8mm 8mm, clip=True, width=0.49\textwidth]{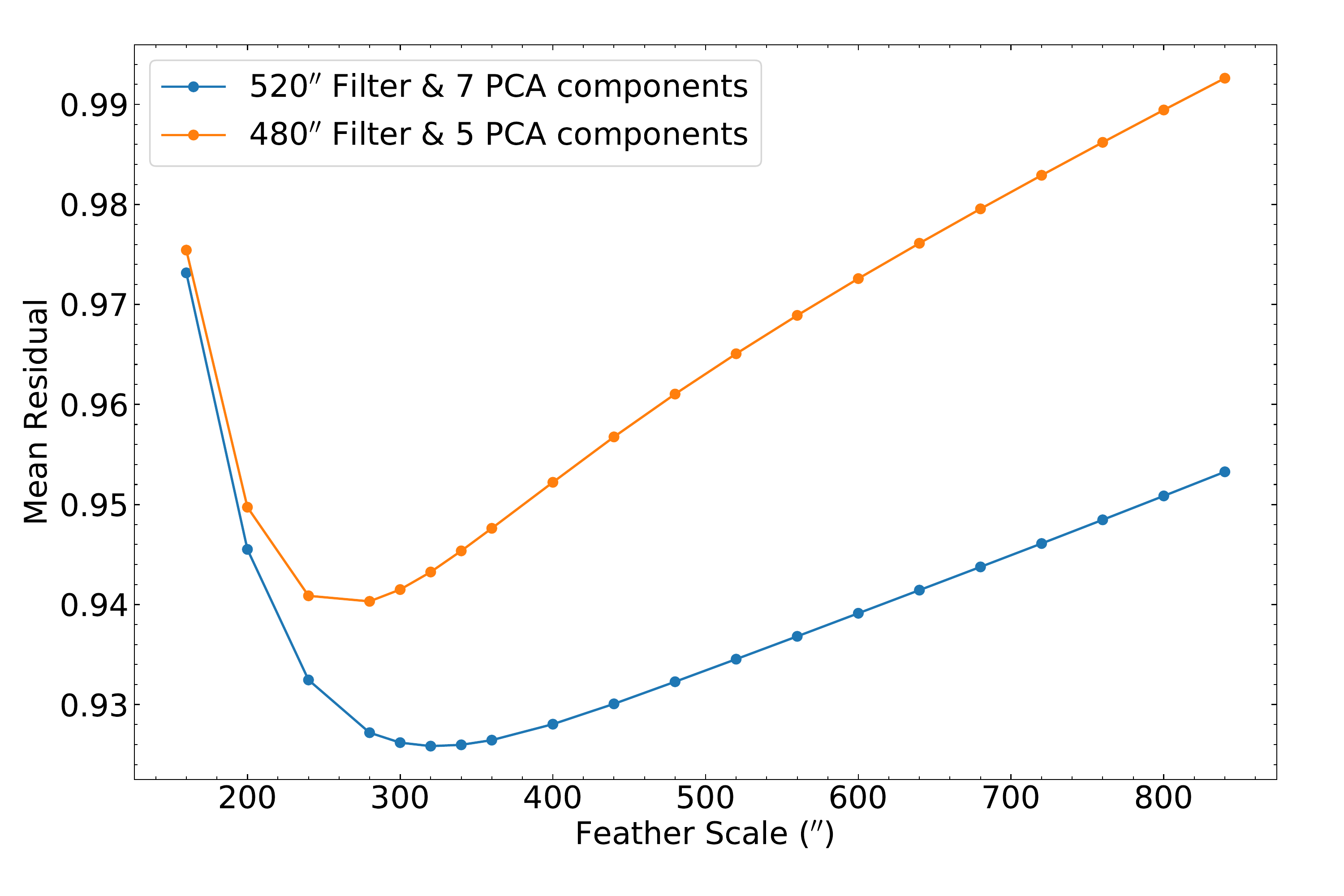}
  \caption{The mean absolute residual in the difference map versus
  the scale of the Gaussian filter, the `feather scale', used
  to combine the SCUBA-2 and {\it Planck} image
  (see the caption of Figure~\ref{fig:meanResidual}
  for details of the difference map and of the region used to measure
  the statistic). The two lines show the results of the simulations
  for our winner and runner-up combinations of high-pass filter scales in the
  SCUBA-2 data-reduction pipeline and number of PCA components.
  The plot shows we get the best results for our winner combination
  (520 arcsec and 7 PCA components) when we combine
  the SCUBA-2 and {\it Planck} images with a Gaussian filter
  with a scale of 320 arcsec.
  }
  \label{fig:featherScalePlot}
\end{figure}

\begin{figure*}
  \centering
  \includegraphics[trim=10mm 25mm 11mm 42mm, clip=True, width=0.98\textwidth]{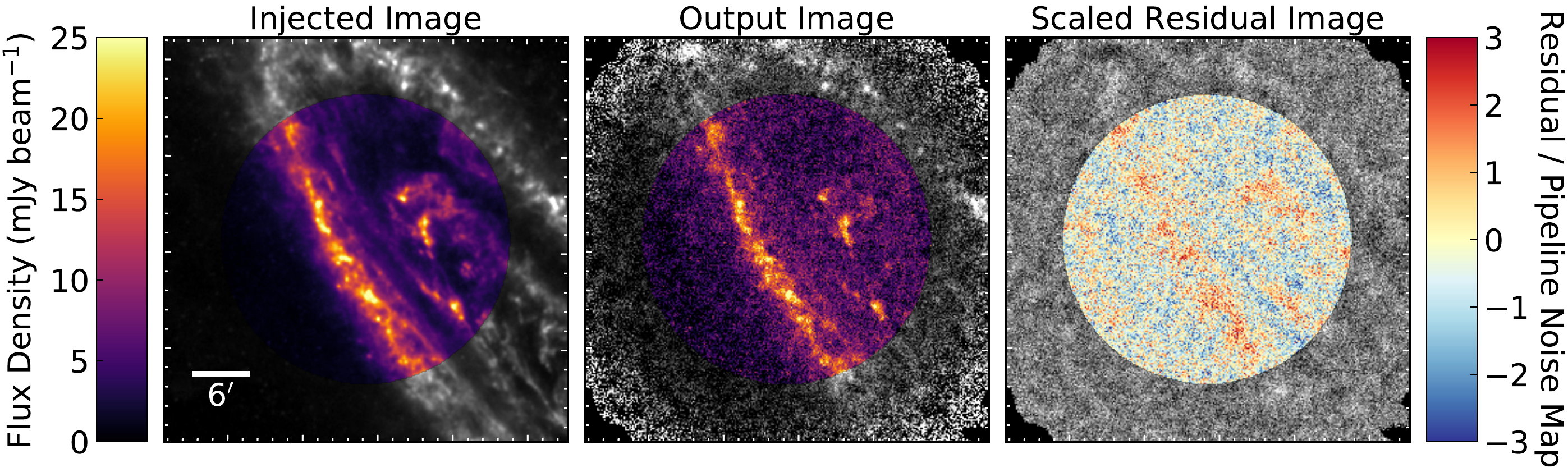}
  \caption{The `true' input image we used in our simulations (left panel),
  the output image we recovered using the data-reduction parameters
  listed in Section~\ref{sec:finalSimParam} (middle panel) and the difference between the
  two divided by the noise estimate from the
  SCUBA-2 data-reduction pipeline (right panel). The
  colored area is the central circular region (diameter
  30 arcmin) in which the observations have their full
  sensitivity.}

  \label{fig:simImages}
\end{figure*}

The final parameter we investigated was the scale of the Gaussian filter,
the `feather scale', used to combine the SCUBA-2 and \textit{Planck} images.
In this final round of simulations, we changed the
pixel scale from 4.0 arcsec to 4.5 arcsec, which makes the
pixels in the final \SI{850}{\micro\meter} image close to one third of the full-width
half maximum of the point spread function, the value that was eventually
adopted after some experimentation for making {\it Herschel} images.
Figure~\ref{fig:featherScalePlot} shows the results from the simulations.
The agreement between the `true' input image and the recovered
output image is best when the scale of the filter is 320\arcsec, which is
roughly what we expected; the resolution of 
\textit{Planck} is 4.8\arcmin\ (290 arcsec) \citep{Planck2014hfi} and so the {\it Planck} image
should supply all Fourier components on angular scales larger
than this. In all the \SI{850}{\micro\meter} simulations the feathering method where the
low-resolution image is not weighted is preferred (see Section~\ref{sec:feather}).

\subsection{The final values of the parameters}
\label{sec:finalSimParam}

As the result of these simulations, we adopted the following values
for all of the parameters when reducing the real SCUBA-2 data:

\begin{itemize}
  \item We set the scale of the
  high-pass filter in the
  SCUBA-2 data-reduction pipeline to
  520\arcsec\ (\textsc{flt.filt\_edge\_largescale}=520).

  \item We set the number of PCA components per array in the SCUBA-2
  data-reduction pipeline to 7 (\textsc{pca.pcathresh}=-7)
  with the values of all the other parameters in the PCA analysis having
  their default values.
  
  \item We use an input mask to define the region where there is likely to be
  astronomical signal created from the {\it Herschel} \SI{500}{\micro\meter} image \citep{smith2012}. This is used by the SCUBA-2 data-reduction pipeline (the AST model in
  the pipeline).
  We defined the mask as all pixels with a \SI{500}{\micro\meter} flux
  greater than
  280\,mJy\,beam$^{-1}$.
  
  \item We set the tolerance parameter in the SCUBA-2 data-reduction pipeline
  to 0.03.
  \item We use a pixel scale for the final \SI{850}{\micro\meter} image of 4.5 arcsec.
  
  \item We used a Gaussian filter with a scale (the `feathering scale') of 320\arcsec\ to combine the final SCUBA-2 image with the {\it Planck} image.
\end{itemize}

These pipeline parameters are optimized for the observing strategy, source properties, and weather for HASHTAG and M\,31. For other
SCUBA-2 datasets with extended structure, we would recommend performing a similar simulation to optimize the processing, however, these results should provide a useful initial guess.  

\subsection{A test of the image fidelity}
\label{sec:faithful}

The final stage in the simulations was to assess the fidelity of the final
image produced with the values of the parameters listed in the previous section.
How close is the structure in the final image to the structure in the original
true image? This analysis gives us a useful estimate of the fidelity of our real image. Note, however, that analysis will yield an upper limit to the errors on
the real image because the simulations have been carried out for
only one {\it Pong} (Section~\ref{sec:surveyStragey}); the spatial overlaps of the many {\it Pong}
fields for the real observations (Figure~\ref{fig:surveyStrategy}) should improve the fidelity
of the final real image.

Figure~\ref{fig:simImages} shows the input `true' image, the recovered output
image, and the difference between the two divided by the noise image
produced by the SCUBA-2 data-reduction pipeline. If our recovery method
was perfect, the final panel should simply show random noise, but in fact there
is some faint structure in the noise that is clearly correlated with bright structures. We therefore need to assess the importance of these systematic
errors.

\begin{figure}
  \centering
  \includegraphics[trim=15mm 20mm 40mm 7mm, clip=True, width=0.47\textwidth]{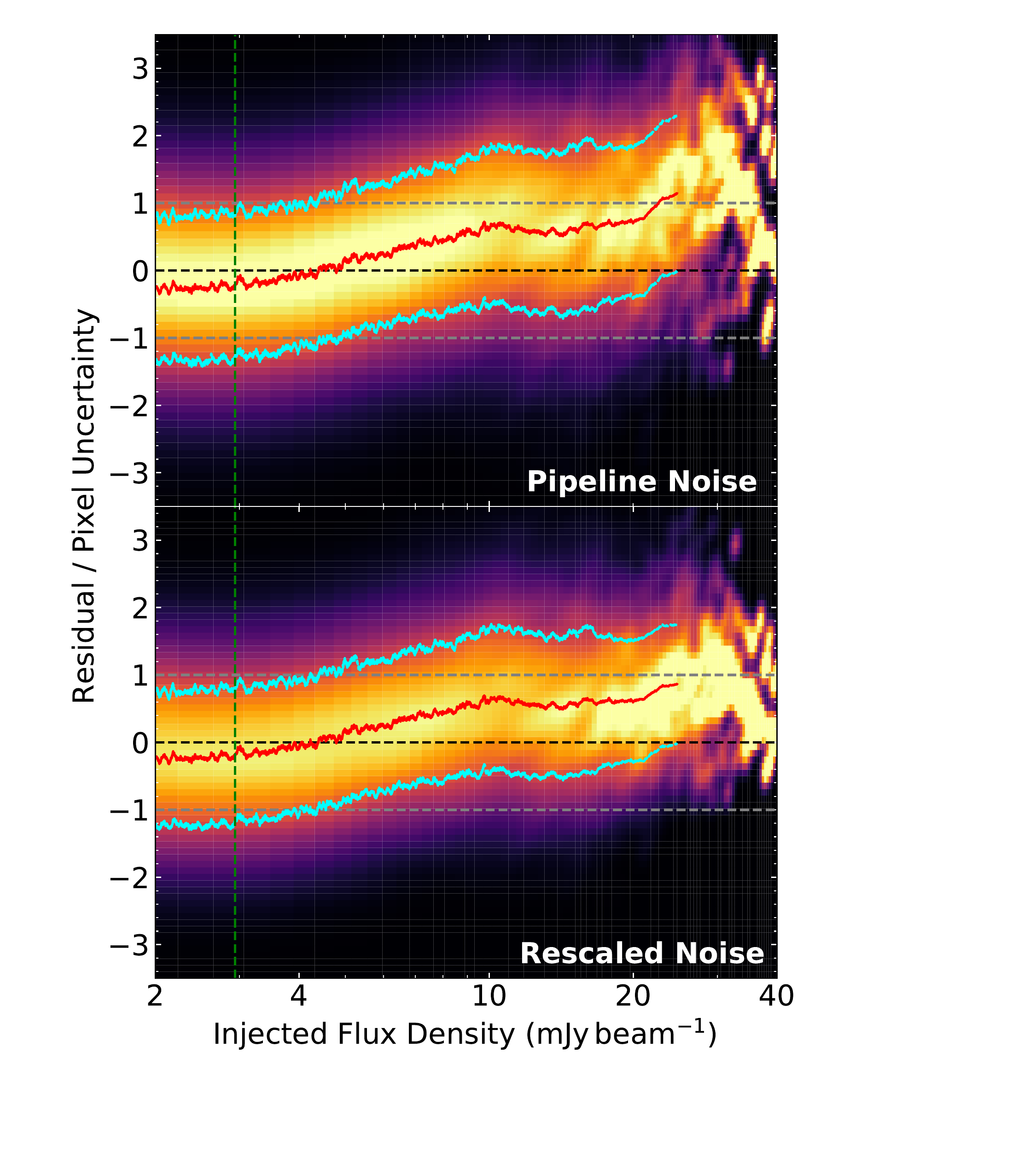}
  \caption{A surface-density plot showing how the number of pixels depends on
  $D$ and the input `true' flux in a pixel. $D$ is the difference between the
  flux density in a pixel in the recovered image and the  flux density in the true image divided by the noise in that pixel. The distribution has been renormalised, column by column, so that the figure is not dominated by the number of pixels
  with faint flux densities. The top panel shows the distribution if the noise that is used is the noise produced by the SCUBA-2 pipeline. 
  The red and cyan lines are the rolling (800 pixels) mean and $\pm1\sigma$ 
  standard deviation, respectively. For a perfect observation
  the red line would follow a mean of zero (black line) and the standard deviation lines would follow the grey lines at $\pm 1$. The green vertical lines show the average 1$\sigma$ 
  noise in the true image. The bottom panel shows the same distribution when the noise has been rescaled using the method described in the text.} 
  \label{fig:noiseScaling}
\end{figure}

We estimated the random and systematic errors in the flux densities
from a plot of $D = (F_{\rm o} - F_{\rm i})/\sigma_{\rm pipe}$ versus $F_{\rm i}$, in which
$F_{\rm o}$ is the flux in a pixel in the output image, $F_{\rm i}$ is the flux in
that pixel in the input image, and $\sigma_{\rm pipe}$ is the estimate
from the SCUBA-2 data-reduction pipeline for the noise in that pixel.
The top panel of Figure~\ref{fig:noiseScaling} is a surface-density plot showing how
the number of pixels depends on $D$ and $F_{\rm i}$. We have only included pixels
in the full-sensitivity 
central circular region of the 
image (diameter of 30 arcmin). 
If the fidelity of the final image was perfect, and if our noise estimates were
correct, $D$ should have a Gaussian distribution
around zero with a standard deviation of one. In reality, the plot shows that there is a clear bias in $D$, which is systematically higher than zero at high flux
densities in the input true image. The standard deviation of $D$ is also slightly higher than one, showing the estimate of the noise produced
by the pipeline is too low.

Given the discrepancy between the actual distribution of $D$ and
its predicted distribution, we have assumed that
the true uncertainty
of the flux in each pixel is given by three uncertainties added in
quadrature:
\begin{equation}
\sigma_{\rm tot}^{2} = \left(a\right)^{2} + \left(b \,\sigma_{\rm pipe}\right)^2 + \left(f\,S_{\rm o}\right)^{2}
\label{equ:totNoise}
\end{equation}
in which $a$ is a constant, $b$ is a multiplicative factor
for the error given by the SCUBA-2 data-reduction pipeline
($\sigma_{\rm pipe}$), and $f$ is a multiplicative
factor for the flux in the output image ($S_{\rm o}$). 
The third error in Equation~\ref{equ:totNoise} is effectively a photometric
calibration error, which exists for all astronomical observations,
although in this case it is an error on top 
of the standard SCUBA-2 photometric calibration error,
which we have not included in the equation.
We estimated the values
of $a$, $b$ and $f$ by applying the 
minimization package
\textsc{lmfit} \citep{lmfit} so that the standard deviation of $D$ 
approached as closely as possible a value of 1. We carried out the minimisation
on a rolling group of 800 pixels ranked in flux.
We found $a$ = 0,
$b = 1.04$, showing the SCUBA-2 data-reduction pipeline had slightly
underestimated the random errors in the fluxes, and
that $f=0.12$, showing that there is a systematic error that depends
on the brightness of the emission, confirming the qualitative impression
produced by Figure~\ref{fig:simImages}. 

The bottom-panel of Figure~\ref{fig:noiseScaling} shows the same
surface-density plot of pixels as in the top panel but with the noise
value predicted by the pipeline ($\sigma_{\rm pipe}$) replaced by the noise value
calculated from Equation~\ref{equ:totNoise} ($\sigma_{\rm tot}$). The distribution shows that while the
width of the distribution of $D$ now is roughly correct, the distribution is still distorted, in the sense that as the output flux density ($F_{\rm o}$) increases, the
output flux density becomes progressively higher than the input flux density
($F_{\rm i}$), although there is a suggestion above 30\,mJy\,beam$^{-1}$ this reduces. We could have corrected the flux densities in the real HASHTAG image
using the red curve in the bottom panel of the figure. 
We decided not to do this for two reasons. First, the real HASHTAG image
is made of a large number of spatially overlapping datasets, so it is possible
this effect is less for the real image. Second, this systematic effect is fairly small compared with the statistical error: only 0.6\,$\sigma$ at $F_{\rm o} = 21$\,mJy\,beam$^{-1}$.

\section{The Simulations at \SI{450}{\micro\meter}}
\label{sec:450sim}

Optimising our data-reduction was much simpler at \SI{450}{\micro\meter} than at \SI{850}{\micro\meter}
because at the shorter wavelength the space-based image
has structure down to a much smaller angular scale ({\it Herschel} --- 36 arcsec) than
at \SI{850}{\micro\meter} ({\it Planck} --- 5 arcmin). The ranges of Fourier components
of the space-based and SCUBA-2 observations are therefore much closer than at the
long wavelength. Nevertheless, we performed the same set of simulations as
at \SI{850}{\micro\meter}, and we summarize the results in this section.

\begin{figure}
  \centering
  \includegraphics[trim=6mm 6mm 8mm 8mm, clip=True, width=0.49\textwidth]{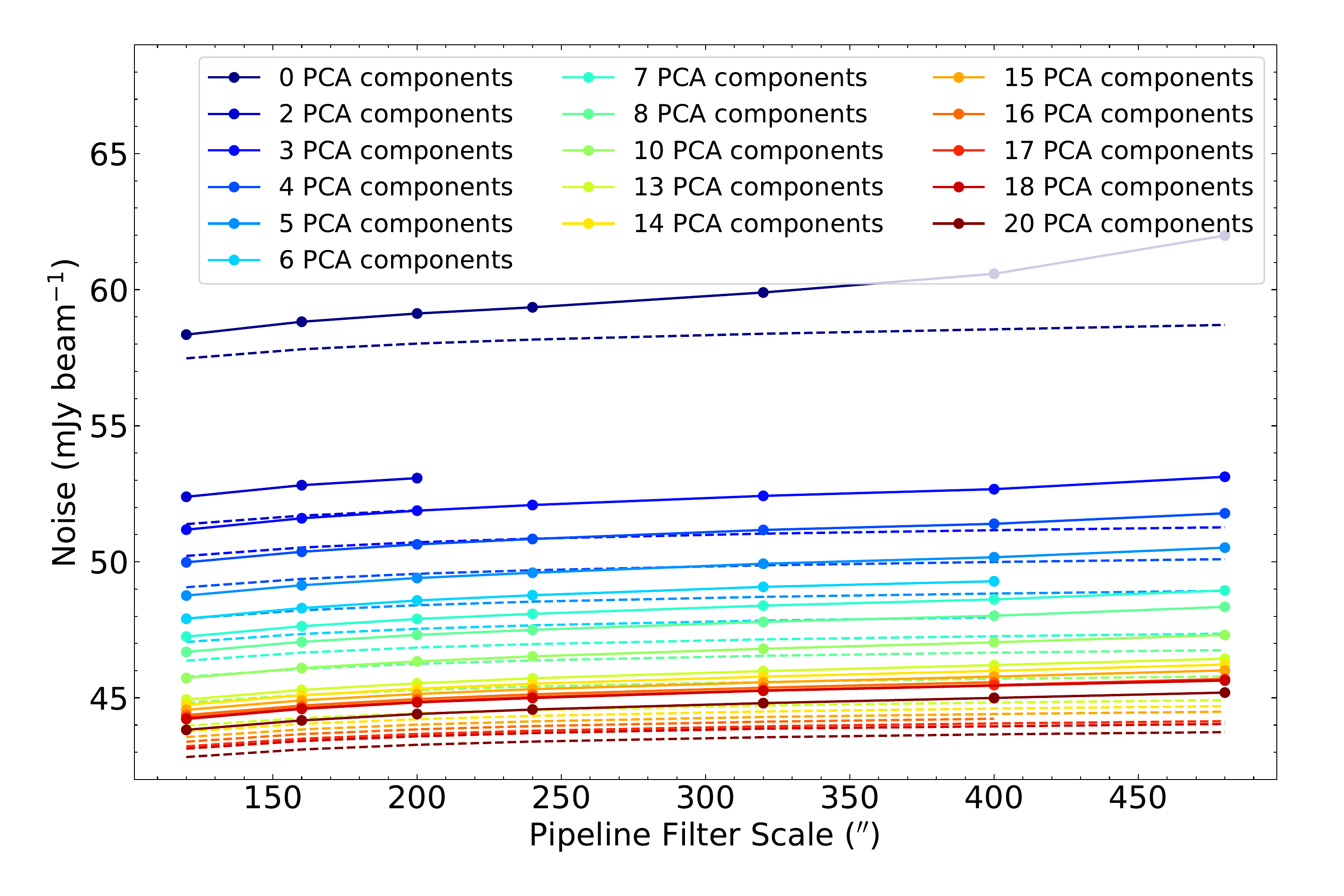}
  \caption{The same plot as Figure~\ref{fig:noisePlot}, but for \SI{450}{\micro\meter}
  data rather than at \SI{850}{\micro\meter}. Increasing the number of PCA components 
  leads to a significant reduction in the noise measured in the image.
  }
  \label{fig:noisePlot450}
\end{figure}

\begin{figure}
  \centering
  \includegraphics[trim=5mm 6mm 8mm 8mm, clip=True, width=0.49\textwidth]{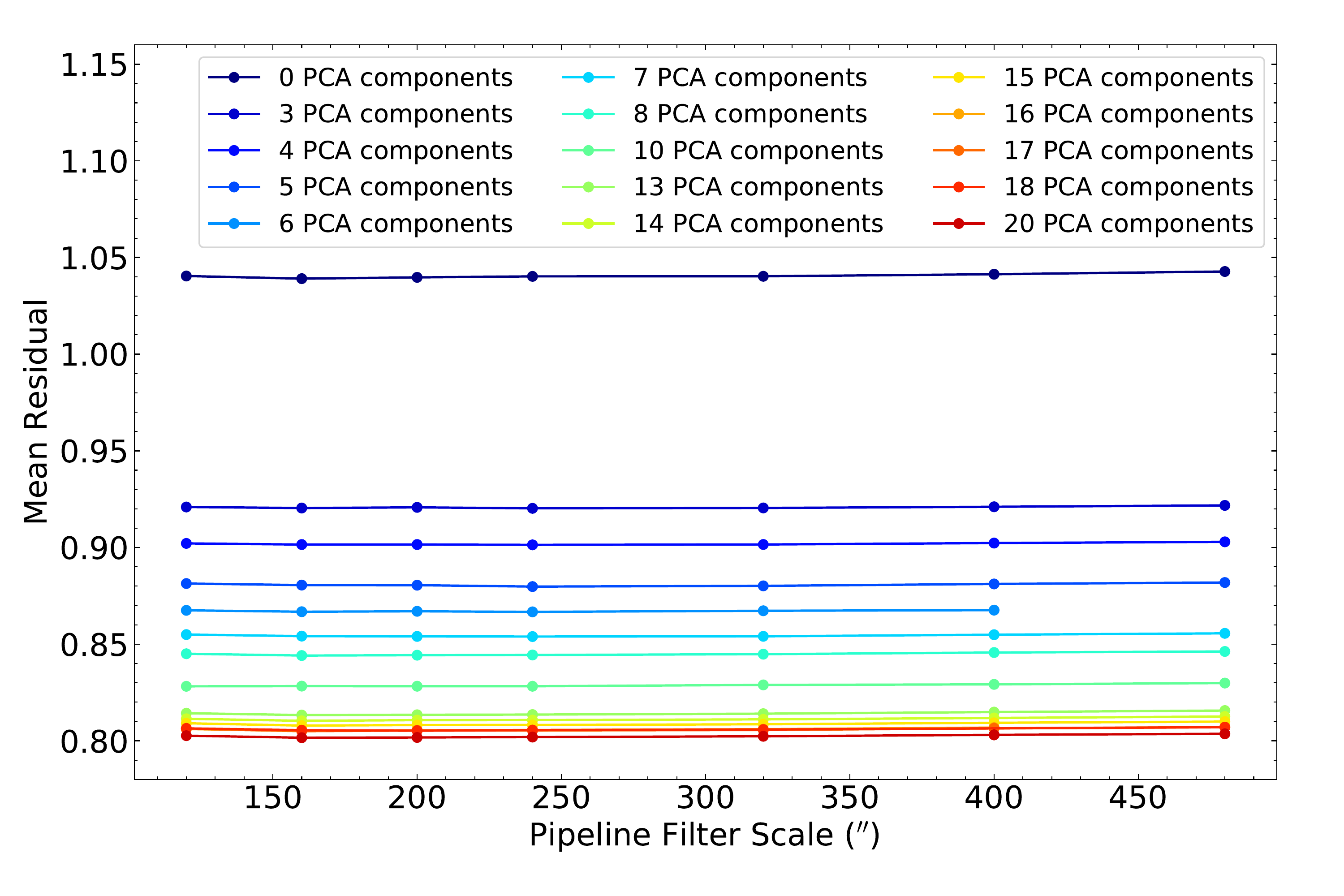}
  \caption{The same plot as Figure~\ref{fig:meanResidual}, but for \SI{450}{\micro\meter}
  data rather than at \SI{850}{\micro\meter}. See the caption of the earlier
  figure for details.
  }
  \label{fig:meanResidual450}
\end{figure}

\begin{figure*}
  \centering
  \includegraphics[trim=8mm 25mm 11mm 42mm, clip=True, width=0.98\textwidth]{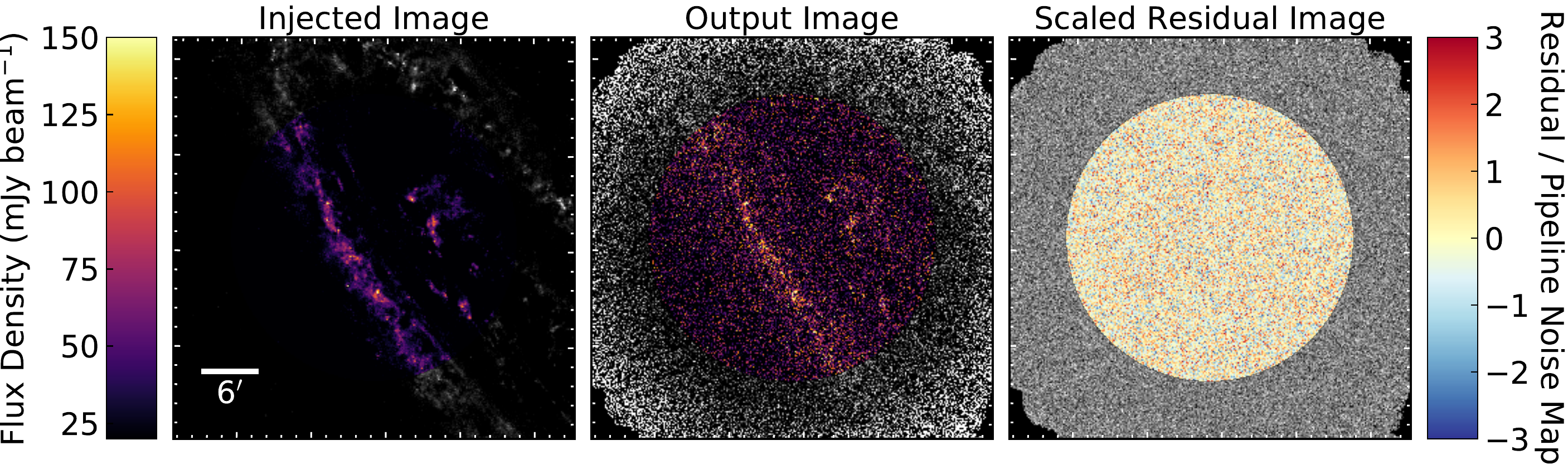}
  \caption{As for Figure~\ref{fig:simImages}, but for the \SI{450}{\micro\meter} simulations.}
  \label{fig:simImages450}
\end{figure*}

Figure~\ref{fig:noisePlot450} shows how the noise in the \SI{450}{\micro\meter} image varies with the filter-scale
and the number of principal components used in the reduction. The number of PCA components is found to be 
particularly important with the noise changing from $\sim$60 to $\sim$45\,mJy\,beam$^{-1}$ with increasing 
number of PCA components.
As before, we investigated the effect on the mean absolute residual in the difference
map of changing the filter scale in the pipeline and the number of PCA
components. We tried filter scales between 120\arcsec\ and 480\arcsec,
with the lower bound chosen because it is used by the 
Cosmology Legacy Survey \citep{Geach2013} which was optimized for detecting point sources. The number of PCA components was again from 0 to 20. Figure~\ref{fig:meanResidual450} shows that the filter scale has little effect
on the mean absolute residual, but that it is reduced by increasing the
number of PCA components. We decided to use a filter scale of 320\arcsec\ with 14 PCA components; there was little improvement by increasing the number of PCA components
further and there was a marginal sign that using a filter with a smaller angular
scale increased the mean absolute residual.

As before, we tested the effect of varying the \SI{500}{\micro\meter} threshold 
used to make the mask and of varying the tolerance value, both of which are
used in the SCUBA-2 pipeline.
We found that the mean absolute residual in the difference map varied very
little when either parameter was adjusted. We therefore decided to use
the same values as at \SI{850}{\micro\meter}.

We tested the effect of varying the feather scale, finding that the optimum
feather scale was 40 arcsec, similar to the size of the {\it Herschel} beam
at \SI{500}{\micro\meter}. We found very little difference 
in the resulting image when using feathering scales up to 100 arcsec (above 50 arcsec
the low-resolution data must be weighted in the feathering, see Section~\ref{sec:feather}).

The final stage in the simulations was to test the fidelity of the image made with
the values of the parameters above. We used exactly the same procedure
as at \SI{850}{\micro\meter} (Section~\ref{sec:faithful}). We found that at \SI{450}{\micro\meter} the noise
estimate from the pipeline ($\sigma_{\rm pipe}$) was a much better estimate of the
true noise ($\sigma_{\rm tot}$) than at \SI{850}{\micro\meter}. We found that the noise scaling term, $b$ in  Equation~\ref{equ:totNoise}, was only 1.02, and the other two terms ($a$ and $f$) were both zero. The input and the output images and the residual map
divided by the noise value from the pipeline are shown in Figure~\ref{fig:simImages450}.
The facts that the residual map shows no structure at all demonstrates that at
\SI{450}{\micro\meter} the random errors are much greater than the systematic errors.

It is known \textit{Herschel} can miss emission on the very largest scales, predominately due to the finite size of the images.
\citet{Clark2021} have investigated this for the Local Group, and found in the case of M\,31 that very little extended dust emission is missing from the SPIRE \SI{500}{\micro\meter} image. We therefore have chosen to feather with the normal SPIRE map, but have provided all the tools and instructions on our website\textsuperscript{\ref{foot:web}} if users wish to feather with an alternative map.

\section{The Real Data}
\label{sec:realData}

\subsection{Calibration}

The SCUBA-2 \textsc{Makemap} routine produces maps in instrumental units of picowatts, and so a flux conversion factor (FCF) is 
used to convert these units into either Jy\,beam$^{-1}$ or Jy\,arcsec$^{-2}$.
There have recently been two changes at the JCMT which has adjusted the standard FCF values used
at the observatory. First, in November 2016 a filter set in SCUBA-2 was changed, which predominantly affected the
\SI{850}{\micro\meter} FCF. Second, in May 2018 the maintenance of the secondary
mirror resulted in a change in the value of the FCF for \SI{450}{\micro\meter}. We
have observations both before and after these changes. Based on an interim analysis by observatory staff (private communication), 
we assume FCF values of 3.62 and 2.14 $\rm Jy\ pW^{-1}\ arcsec^{-2}$ at 450 and \SI{850}{\micro\meter}, respectively, for observations post May 2018\footnote{Since we calibrated the data, new (but still preliminary) values of the FCF
have been released \url{https://www.eaobservatory.org/jcmt/instrumentation/continuum/scuba-2/calibration/}, which are slightly different 
from the values we use (resulting in approximately 3.5\% and 5\% lower flux-density at 450 and \SI{850}{\micro\meter}, respectively). As the correction depends on 
individual observing conditions, we will apply the new calibration in the next data release.}.
For our older observations, we multiply the `cleaned' data (see Section~\ref{sec:initialProcessing}) by a correction factor, so that the final data can be calibrated with the same FCF. We adopted values for this correction factor of 1.21053 and 1.06481, for 450 and \SI{850}{\micro\meter}, respectively.

The final factor we have to consider is that the standard JCMT calibration scheme
is not designed for such extended objects as Andromeda.
The JCMT calibration scheme \citep{Dempsey2013} uses 
the flux of a calibrator source within a 30\arcsec\ radius aperture,
after subtracting a background measured in an annulus around the calibrator between
radii of 45 and 60 arcsec. This scheme is fine for calibrating images that
contain point sources. But we are trying to calibrate very extended emission, and
the beam of the telescope extends to much larger radii than the radii used in
the standard calibration scheme.

We have adopted the calibration scheme we devised for the JINGLE Large Programme \citep{Smith2019}, where we multiply the FCFs so that the flux densities in the images matched the convention of other telescopes, that an aperture centered on a galaxy would only include the entire flux of the galaxy if the radius was increased
to infinity. By integrating the beam model in \citet{Dempsey2013},
we calculated that the standard \SI{850}{\micro\meter} FCF should be multiplied by 0.91, which agreed with the curve of growth found by \citet{Dempsey2013}. 
Doing the same calculation at \SI{450}{\micro\meter}, we obtain a correction factor of 0.99. The curve of growth given in \citet{Dempsey2013}, however,
suggests a slightly smaller factor of 0.97. We suspect the
difference is caused by extra non-Gaussian features in the beam at large radii.
We therefore adopt the smaller value of 0.97. The final values of the FCF used to 
create these HASHTAG images were therefore the FCF values given above multiplied
by these correction factors. These are 3.51 and 1.95\,Jy\,pW$^{-1}$\,arcsec$^{-2}$ at 450 and \SI{850}{\micro\meter}, respectively.

\subsection{Final Maps}
\label{sec:finalMaps}

The real HASHTAG data was reduced using the method outlined in the previous
subsections, using the best values of the parameters found from the simulations.
Our final maps (Data Release 1, DR1) are composed of two multi-extension fits files which contain the flux-density, uncertainty, and sensitivity maps for the 
450 and \SI{850}{\micro\meter} images, respectively. 

The uncertainty map contains the true uncertainty values we derived
in Section~\ref{sec:faithful}, which includes both the statistical uncertainty in
each pixel and the systematic uncertainty as the result of the flux density
in that pixel (the third term in Equation~\ref{equ:totNoise}). The sensitivity map is the same
except that this map does not include the systematic term ($f$ in Equation~\ref{equ:totNoise} is set to zero). At both wavelengths, the sensitivity of our final images exceeds our
targets. In the 10\,kpc ring, the typical sensitivity is $\sim$2.0 and $\sim$30\,mJy\,beam$^{-1}$ at 850 and \SI{450}{\micro\meter}, respectively, with peak
sensitivities in the center of 1.5 and 20.6\,mJy\,beam$^{-1}$ at 850 and \SI{450}{\micro\meter},
respectively. As a rough comparison the \textit{Herschel} \SI{500}{\micro\meter} observations have an instrumental sensitivity of $\sim$11\,mJy\,beam$^{-1}$ \citep{Smith2017},
and the point source sensitivity of \textit{Planck} at \SI{850}{\micro\meter} is $\sim$69\,mJy\,beam$^{-1}$ \citep{Planck2014b}.

\begin{figure*}
  \centering
  \includegraphics[trim=1mm 0mm 0mm 3mm, clip=True, width=0.75\textwidth]{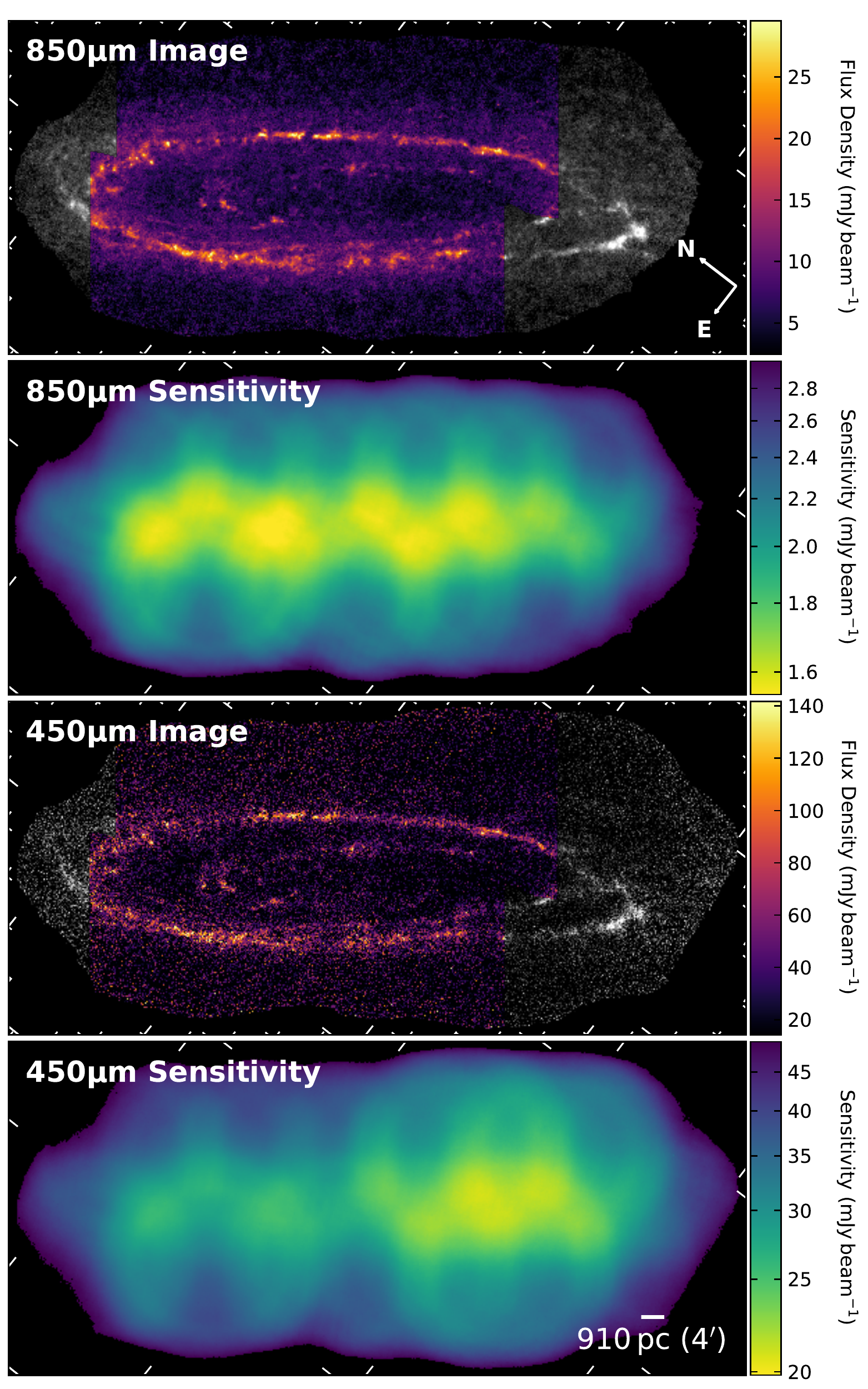}
  \caption{Our final 450 and \SI{850}{\micro\meter} images and sensitivity maps for HASHTAG Data Release~1. 
           The colored regions of the images are approximately where our observations and complete and at our
           full sensitivity (equivalently, the grey-scale shows regions where observations are still ongoing). The images
           have a resolution of 7.9\arcsec\ and 13.0\arcsec (FWHM), at 450 and \SI{850}{\micro\meter}, respectively, and are available 
           in both mJy\,beam$^{-1}$ and mJy\,arcsec$^{-2}$ units.
           The sensitivity maps are shown on a log scale and are described
           in Section~\ref{sec:finalMaps}.}
  \label{fig:DR1images}
\end{figure*}

The flux-density and sensitivity maps are shown in Figure~\ref{fig:DR1images}. 
The sensitivity is not quite as uniform at \SI{450}{\micro\meter}, which we attribute to variations in sub-mm opacity during the periods we took the data, since 
opacity variations have a bigger effect at the shorter wavelength.

As well as the images shown in Figure~\ref{fig:DR1images}, we also provide versions that have three different levels of Gaussian smoothing 
(for example, Figure~\ref{fig:smoDR1images}), so that users can balance resolution versus signal-to-noise ratio. In these smoother images, the raw images have been
smoothed with a Gaussian with a FWHM of
4, 5 and 7.9\arcsec\ at \SI{450}{\micro\meter} and  7, 10 and 13\arcsec\ at \SI{850}{\micro\meter} (this equates to effective resolutions of
8.9, 9.3, and 11.2\arcsec\ at \SI{450}{\micro\meter}, and 15.2, 16.8, and 19.1\arcsec\ at \SI{850}{\micro\meter}). 

\section{CO($J$=3--2) Subtraction}
\label{sec:CO}

A possible contamination in the \SI{850}{\micro\meter} image is the contribution of the CO($J$=3--2) line which has a rest frequency of 345.796\,GHz, putting it within the \SI{850}{\micro\meter} passband. This line has been shown to contribute anything from 0.7\% to 41\% of the \SI{850}{\micro\meter} emission 
in nearby galaxies, although it is normally less than 15\% \citep{Smith2019}, and in the Milky Way the contamination
is typically small ($<$5\%) but can be 30\% \citep{Moore2015}. The contamination seems likely to be low in M\,31 because of the low fraction of molecular gas. We can get a rough idea of the likely
scale of the contamination using our CO($J$=3--2) survey of selected
regions within the galaxy \citep{Li2020}. In our
CO survey the strongest line flux we found was
$I_{\rm CO(\textit{J}=3-2)} \simeq 5\,{\rm K\,km\,s^{-1}}$, which
corresponds, using the relationship given in \citet{Parsons2018}, to an \SI{850}{\micro\meter} flux density
of $\simeq$3 mJy\,beam$^{-1}$, $\simeq$1.5 times the typical noise
(Section~\ref{sec:finalMaps}).

We used the results of our CO($J$=3--2) survey, which covered
small regions but over a range of environments (Figure~\ref{fig:surveyStrategy}), as the ground truth for developing a method to estimate
the CO contamination at all points across M\,31. As our starting point, we used
the CO($J$=3--2) cubes presented by \citet{Li2020}.
However, we created a new set of integrated intensity maps (moment-0 maps), using a
new VLA H{\sc i} dataset \citep{Koch2021}. The VLA dataset provides a 58\arcsec\ resolution (FWHM) image of the entire galaxy, and a higher resolution
18\arcsec\ image of the region covered by \textit{Hubble}. To create the CO integrated intensity maps we use the H{\sc i} data as a prior to mask the channels
in the cube that are not expected to contain any CO emission, using the moment-1 map to predict the center of the line with a width based on the H{\sc i} line-width  (with a minimum 40\,km\,s$^{-1}$ adopted). Using the H{\sc i} as a prior is similar technique to that applied by \citet{Schruba2011}. 

Our overall approach was to use a combination of
several datasets to predict the CO($J$=3--2) emission, using our own CO($J$=3--2) survey to determine the combination that gives the best prediction. 
We included maps of the
dust emission \citep[column-density, temperature, and emissivity index;][]{Whitworth2019}, images in the WISE bands \citep{Wright2010,Cutri2013}, 
in the UV \citep{Thilker2005}, in the mid-infrared \citep[{\it Spitzer}, MIPS;][]{gordon2006}, a survey of part
of the galaxy with CARMA in CO($J$=1--0) \citep[][A.~Schruba et al., \textit{in preparation}]{CalduPrimo2016}, and a map of the
estimated star-formation rate in the galaxy calculated using UV and \SI{24}{\micro\meter} \citep{ford2013}. 
While some of these datasets may be degenerate, our aim was to find the best model to predict the CO($J$=3--2) line flux, rather than understanding
the physical meaning of the obtained model.

\begin{figure}
  \centering
  \includegraphics[trim=1mm 12mm 2mm 46mm, clip=True, width=0.49\textwidth]{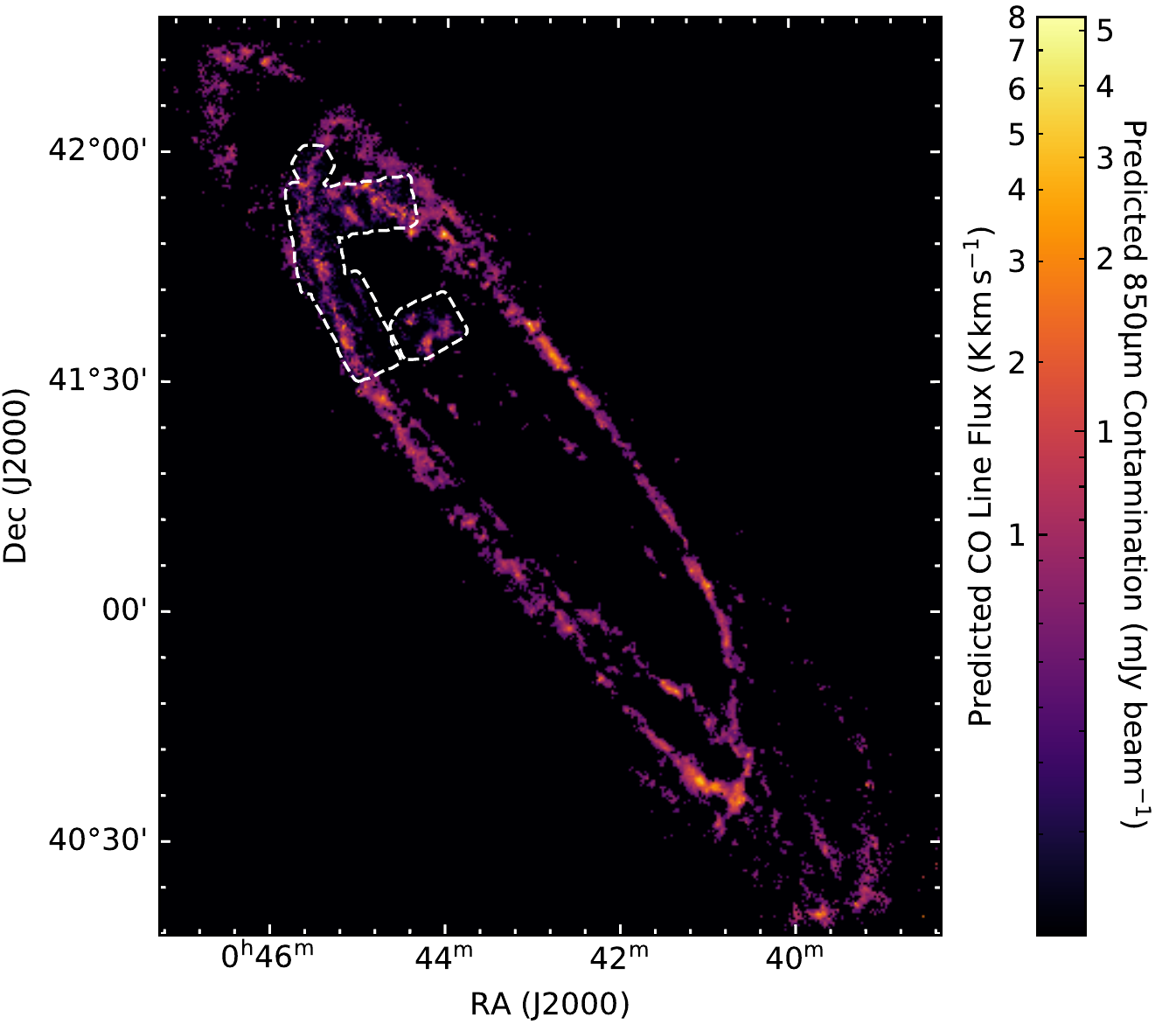}
  \caption{Our map of the CO line predicted (or equivalently \SI{850}{\micro\meter} continuum contamination)
  using the method described in Section~\ref{sec:CO}. Outside
  the CARMA region, any pixels with predicted
  line fluxes $<$0.5\,K\,km\,s$^{-1}$ have been set to
  zero, because these pixels fall outside
  the range of CO line fluxes where the model
  was trained. The peak predicted line flux is $\sim$15\,K\,km\,s$^{-1}$, but we have capped
  the color-bar at 8\,K\,km\,s$^{-1}$ to aid visibility. The white-dashed line shows the region covered by
  CARMA CO($J$=1--0) observations.}
  \label{fig:COmap}
\end{figure}

The most useful dataset for predicting the CO($J$=3--2) emission is a map in another CO line.
We decided not to use the map in the CO($J$=1--0) line 
over the whole galaxy \citep{Nieten2006} because its
resolution (23\arcsec) 
is significantly lower than that of our  \SI{850}{\micro\meter} image (13\arcsec), and we found a model derived at 23\arcsec\ resolution 
when applied on 13\arcsec\ scales did not perform as well as one derived at the higher resolution.
The CARMA survey was our key dataset
because it was a survey in the CO($J$=1--0)
line with a resolution of 5.5\arcsec\ (we use the version corrected for missing large-scale CO emission using the IRAM single-dish data). 
The survey, however, only covers $\sim$323\,arcmin$^2$ (see Figure~\ref{fig:COmap}) and
our CO($J$=3--2) survey covers an even smaller region (see Figure~\ref{fig:surveyStrategy}). We were
therefore forced to develop a two-stage method.

In the first stage of the approach, we restricted
our analysis to the small portion of the galaxy
covered by our CO($J$=3--2) survey that was also within
the region covered by CARMA. We do not explicitly include the CO($J$=3--2)/CO($J$=1--0) ratio which has been found to vary across M\,31 \citep{Li2020}, but
implicitly include it in the model. 
We first performed a background subtraction on the input continuum images where necessary, convolved these images to the same resolution as the \SI{850}{\micro\meter}/JCMT CO products, and finally re-projected them to match each of the six JCMT CO($J$=3--2) integrated intensity maps that overlap with the CARMA field. We took as our inputs the logarithms
of all the input maps except for the dust temperature and 
emissivity index. We assigned the pixels randomly
into a training set (80\% of the data) and
a testing set (20\% of the data). We then used the \textsc{scikit-learn} \citep{scikit-learn} `standard scaler' routine which standardizes each of the inputs by removing the mean and dividing by the standard deviation. To perform the fitting, we tried both linear and non-linear methods (random forest, 
and the Multi-layer Perception neural network) but found the linear model performed as well as the more complicated routines. As the JCMT data is relatively noisy, to incorporate the uncertainties we built a model using \textsc{pymc3} \citep{pymc3}, in which the predicted CO value is given by:
\begin{equation}
y_{\rm model} = 10^{c\cdot \sum_i m_i x_i }
\label{equ:COmodel}
\end{equation}
in which $c$ is a constant, $m_i$ is the gradient of each input, and $x_i$ is each input `feature' (i.e., each input image). For both the constant and gradients we
assumed a weakly-informative Gaussian prior with $\mu = 0$ and $\sigma=10$ for the intercept and $\sigma = 20$ for the gradient.
Table~\ref{tab:COparam} provides the best-fitting values of $m_i$ and and $c$.

\begin{deluxetable}{ccc}
\tablecaption{Parameters of CO($J$=3--2) Models \label{tab:COparam}}
\tablecolumns{3}
\tablewidth{0pt}
\tablehead{`Feature' & \multicolumn{2}{c}{Gradient Coefficient ($m_i$)}\\
           Image & Including CARMA & Excluding CARMA}
\startdata
CARMA & \hspace{7pt}$0.403 \pm 0.020$ & \mdash \\
Dust Surface- & \multirow{2}{*}{\hspace{7pt}$0.076 \pm 0.023$} & \multirow{2}{*}{\hspace{7pt}$0.220 \pm 0.004$}\\
Density & & \\
Dust Temperature  & $\hspace{7pt}0.060 \pm 0.019$ & \hspace{7pt}$0.160 \pm 0.005$\\
Dust $\beta$  & \hspace{7pt}$0.032 \pm 0.009$ & \hspace{7pt}$0.030 \pm 0.003$\\
MIPS \SI{24}{\micro\meter} & \hspace{7pt}$0.014 \pm 0.049$ & \hspace{7pt}$0.133 \pm 0.010$\\
SFR Surface- & \multirow{2}{*}{\hspace{7pt}$0.011 \pm 0.039$} & \multirow{2}{*}{$-0.115 \pm 0.009$}\\
Density & & \\
WISE W1 & $-0.173 \pm 0.053$ & $-0.153 \pm 0.011$\\
WISE W2 & \hspace{7pt}$0.046 \pm 0.058$ & \hspace{7pt}$0.079 \pm 0.012$\\
WISE W3 &\hspace{7pt}$0.030 \pm 0.044$ & $-0.076 \pm 0.008$\\
WISE W4 & \hspace{7pt}$0.022 \pm 0.042$ & \hspace{7pt}$0.063 \pm 0.010$\\
\hline
Constant ($c$) & $-0.288 \pm 0.016$ & \hspace{7pt}$0.420 \pm 0.0003$\\
\enddata
\tablecomments{The best-fit parameters from the model described by Equation~\ref{equ:COmodel}, for both
the model including CARMA observations (`stage 1') or excluding CARMA observations (`stage 2'). See 
Section~\ref{sec:CO} for more details.}

\end{deluxetable}

\begin{figure}
  \centering
  \includegraphics[trim=4mm 4mm 0mm 0mm, clip=True, width=0.49\textwidth]{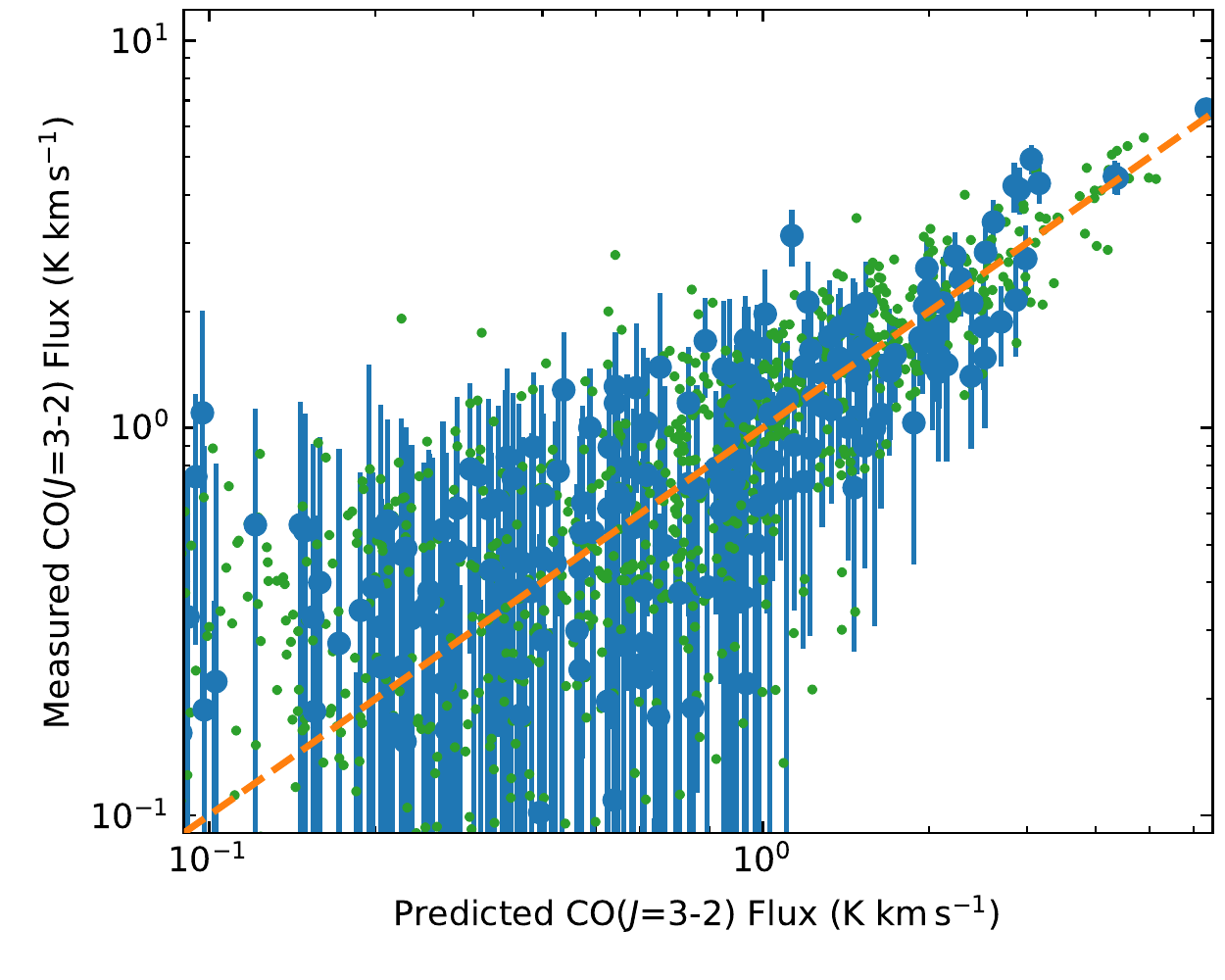}
  \caption{The measured CO($J$=3--2) flux versus the
  flux predicted by our linear model for the
  six regions of our CO($J$=3--2) survey that fall within
  the region of the CARMA($J$=1--0) survey. The blue data points are for the 20\% of pixels that
  are in our test dataset, which we did not use in determining the best combination of parameters. The green points are from our training dataset. At low line-fluxes the scatter is dominated by the uncertainty in the JCMT CO($J$=3--2) observations.
  The orange  line shows the 1-to-1 line that we would achieve if our prediction method was perfect.
  }
  \label{fig:CARMAtraining}
\end{figure}

Figure~\ref{fig:CARMAtraining} shows a plot of the predicted versus the measured CO($J$=3--2) emission. Above 1.0\,K\,km\,s$^{-1}$ the average accuracy is $\sim$28\%,
although the true uncertainty may be lower as there is significant uncertainty associated with some of the data points.

In the second stage of our method, we extend our analysis to create a model for regions of M\,31 where
we do not have CARMA CO($J$=1--0). To create this model we extend our analysis to the entire CARMA region, for which
we have CO($J$=1--0) observations but not CO($J$=3--2)
observations. In this much larger region we use the linear combination we derived in
stage 1 to predict the CO($J$=3--2) fluxes. We then use
these predictions as the `measurements' in this stage
of the analysis, as well as our JCMT CO($J$=3--2) measurements not used in `stage 1' (e.g., outside the CARMA footprint). 
In this stage we use equation~\ref{equ:COmodel} as above to determine the combination of inputs that makes
the best prediction of the CO($J$=3--2) `measurements', except
this time we do not use the CARMA CO($J$=1--0) measurements as one of the inputs.

\begin{figure}
  \centering
  \includegraphics[trim=0mm 4mm 0mm 0mm, clip=True, width=0.49\textwidth]{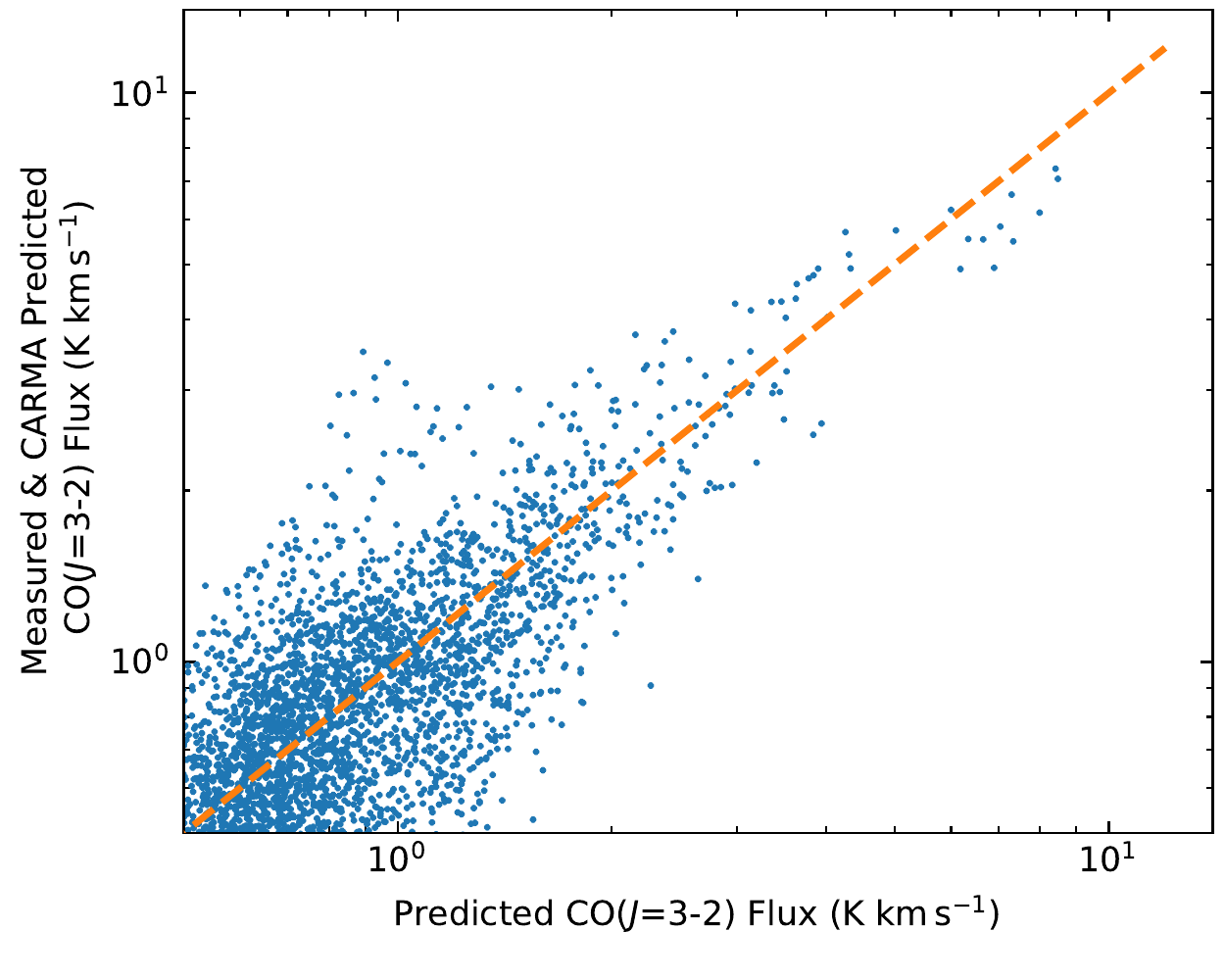}
  \caption{The CO($J$=3--2) line flux predicted by
  our final model versus CO($J$=3--2) measurements (either
  our real measurements or the predictions from stage 1). The blue data points are
  from our `test' dataset (20\%
  of the pixels), which was not used to derive the model. Error bars have not been included
  for clarity. The orange line shows the 1-to-1 line for a perfect prediction.
  }
  \label{fig:COmodel}
\end{figure}

We found, as before, that there was no advantage when using the non-linear methods, so we used the simpler linear
method. CARMA covers a large continuous region, 
so instead of assigning pixels randomly to the
training and test data, we used slices in declination,
which avoids pixels in the same cloud being assigned
to both datasets. Since predicted
CO($J$=3--2) fluxes below 0.5\,K\,km\,s$^{-1}$ correspond
to \SI{850}{\micro\meter} fluxes significantly less than
the statistical noise in the HASHTAG image,
we only trained our model on regions with
CO($J$=3--2) `measurements' greater than this value.
Figure~\ref{fig:COmodel} shows the relationship between
the prediction of this new model versus the CO($J$=3--2)
measurements (either our real measurements or
the predictions from stage 1).

Figure~\ref{fig:COmap} shows the CO($J$=3--2) line flux predicted by our models. Inside the CARMA region, 
where we have CO($J$=1--0) measurements, we have used the stage 1 model. Outside the CARMA region, we used our
stage 2 model. The statistical noise in the
HASHTAG \SI{850}{\micro\meter} image is $\simeq$2\,mJy, which
corresponds to a CO($J$=3--2) line flux of $\simeq3$\,K\,km\,s$^{-1}$. The figure shows that generally the
line contamination is not a problem. In bright cores,
though, it can be important. If $\sigma_{850\mu{\rm m}}$ is the
statistical uncertainty in the \SI{850}{\micro\meter} without
the inclusion of the systematic term (the third term
in equation 1), the maximum CO($J$=3--2) signal is $\simeq5.7\sigma_{850\,{\rm\mu m}}$. But if the systematic
term is included, this reduces to $\simeq 1.5\, \sigma$.
If only pixels
are included where the signal-to-noise ratio of the \SI{850}{\micro\meter} image is greater than 3\,$\sigma$ (not including the systematic term), the maximum
contamination in a pixel is 28\%, but in 80\% of the
pixels the CO line flux is less than
0.5\,K\,km\,s$^{-1}$, which is only $\simeq$16\% of the statistical noise in the continuum map. In general, then, contamination by the CO($J$=3--2) 
line is not a significant problem. We have provided
\SI{850}{\micro\meter} images in the data release both with and
without a correction for line contamination, allowing users to either ignore the effect of
line contamination completely, use our corrected image,
or make their own correction.

\section{Conclusions}

We have presented submillimeter images of the Andromeda Galaxy obtained at 450 and \SI{850}{\micro\meter}, the first images
made from the ground that  properly represent the structure of the galaxy on all
spatial scales. We have described the method we have developed to optimize the SCUBA-2 pipeline for M\,31
and how we use a feathering technique to combine the small-scale structure
(high-\textit{k} Fourier components) from SCUBA-2 and data from
space observatories ({\it Herschel} and {\it Planck}) to
provide the large-scale structure (low-\textit{k} Fourier components).

We describe the maps that comprise the HASHTAG DR1 data release which have a typical sensitivity of $\sim$2.0 and 
$\sim$30\,mJy\,beam$^{-1}$ at 850 and \SI{450}{\micro\meter}, respectively (at native SCUBA-2 resolution).
As the CO($J$=3--2) line falls with the bandpass of the \SI{850}{\micro\meter} band we derive a method to 
predict the line flux across M\,31, and find while generally the contamination is small compared with the uncertainty
in our continuum measurements, for some bright regions of the ring the contamination is significant. We provide
data products both with and without the CO correction, and at different resolutions.


\acknowledgements

The James Clerk Maxwell Telescope is operated by the East Asian Observatory on behalf of The 
National Astronomical Observatory of Japan; Academia Sinica Institute of Astronomy and Astrophysics; 
the Korea Astronomy and Space Science Institute; the Operation, Maintenance and Upgrading Fund for 
Astronomical Telescopes and Facility Instruments, budgeted from the Ministry of Finance (MOF) of 
China and administrated by the Chinese Academy of Sciences (CAS), as well as the National Key R\&D 
Program of China (No. 2017YFA0402700). Additional funding support is provided by the Science and 
Technology Facilities Council of the United Kingdom and participating universities in the United Kingdom and Canada.
Additional funds for the construction of SCUBA-2 were provided by the Canada Foundation for Innovation.

This research was undertaken using the supercomputing facilities at Cardiff University operated by Advanced Research Computing at Cardiff (ARCCA) on behalf of the Cardiff Supercomputing Facility and the HPC Wales and Supercomputing Wales (SCW) projects. We acknowledge the support of the latter, which is part-funded by the European Regional Development Fund (ERDF) via the Welsh Government.

This research made use of Astropy, a community-developed core Python package for Astronomy (Astropy Collaboration, 2013).

MWLS acknowledges funding from the UK Science and Technology Facilities Council 
consolidated grant ST/K000926/1.

MWLS and HLG acknowledges support from the European Research Council (ERC) in the form of Consolidator Grant {\sc CosmicDust} (ERC-2014-CoG-647939). 

BL and AC acknowledge the support from the National Research Foundation grant No. 2018R1D1A1B07048314.

BL acknowledges support from the National Science Foundation of China (12073002, 11721303).

MB was supported by STFC consolidated grant `Astrophysics at Oxford' ST/H002456/1 and ST/K00106X/1.

TAD acknowledges support from STFC grant ST/S00033X/1.

JH thanks the National Natural Science Foundation of China under grant Nos. 11873086 and U1631237 and support by the Yunnan Province of China (No.2017HC018).

This work is sponsored (in part) by the Chinese Academy of Sciences (CAS), through a grant to the CAS South America Center for Astronomy (CASSACA) in Santiago, Chile.

YG was supported by National Key Basic R\&D Program of China (2017YFA0402704),  NSFC Grant 11861131007, 12033004, and Chinese Academy of Sciences Key Research Program of Frontier Sciences (QYZDJ-SSW-SLH008).

LCH was supported by the National Science Foundation of China (11721303, 11991052) and the National Key R\&D Program of China (2016YFA0400702).

T.M.H. acknowledges support from the Chinese Academy of Sciences (CAS) and the National Commission for Scientific and Technological Research of Chile (CONICYT) through a CAS-CONICYT Joint Postdoctoral Fellowship administered by the CAS South America Center for Astronomy (CASSACA) and CONICYT in Santiago, Chile.

FK acknowledges support by Academia Sinica under Investigator Award AS-IA-106-M03, and by the Ministry of Science and Technology (MoST) of Taiwan under grant MOST107-2119-M-001-031-MY3.

FlK was supported by European Research Council Grant SNDUST ERC-2015-AdG-694520.

Z.N.L. and Z.Y.L. acknowledge support by the National Key Research and Development Program of China (2017YFA0402703) and National Natural Science Foundation of China (grant 11873028).

MJM acknowledges the support of the National Science Centre, Poland through the POLONEZ grant 2015/19/P/ST9/04010 and SONATA BIS grant 2018/30/E/ST9/00208.

CDW acknowledges support from the Natural Sciences and Engineering Research Council of Canada and the Canada Research Chairs program

TGW acknowledges funding from the European Research Council (ERC) under the European Union's Horizon 2020 research and innovation programme (grant agreement No. 694343).



\bibliographystyle{aasjournal}
\bibliography{hashtag} 

\end{document}